\documentclass[conference]{IEEEtran}
\usepackage{setspace}
\usepackage{algorithm2e}
\usepackage{epsfig}
\usepackage{graphicx}
\usepackage{wrapfig}
\graphicspath {{Figures/}}\DeclareGraphicsExtensions{.eps,.ps}
\usepackage{comment}
\usepackage{subfigure}
\usepackage{multirow}
\usepackage{amssymb,amsmath}

\small\normalsize
\IEEEoverridecommandlockouts
\begin{document}
\pagestyle{plain}
\title{Link Failure Recovery over Very Large Arbitrary Networks: The Case of Coding}
\author{\IEEEauthorblockN{Serhat Nazim Avci and Ender Ayanoglu}\\
\IEEEauthorblockA{Center for Pervasive Communications and Computing\\
Department of Electrical Engineering and Computer Science\\
University of California, Irvine\\
Irvine, CA 92697-2625
}
\thanks{This work is partially supported by NSF under Grant No. 0917176.}
}
\maketitle
\begin{abstract}
%
\boldmath
Network coding-based link failure recovery techniques provide near-hitless recovery and offer high capacity
efficiency. Diversity coding is the first technique to incorporate coding in this field and is easy to implement
over small arbitrary networks. However, its capacity efficiency is restricted by its systematic coding and high design complexity even though its design complexity is lower than the other coding-based recovery techniques. Alternative techniques mitigate some of these limitations, but they are difficult to implement over arbitrary networks. In this paper, we propose a simple column generation-based design algorithm and a novel advanced diversity coding technique to achieve near-hitless recovery over arbitrary networks. The design framework consists of two parts: a main problem and subproblem. Main problem is realized with Linear Programming (LP) and Integer Linear Programming (ILP), whereas the subproblem can be realized with different methods. The simulation results suggest that both the novel coding structure and the novel design algorithm lead to higher capacity efficiency for near-hitless recovery. The novel design algorithm simplifies the capacity placement problem which enables implementing diversity coding-based techniques on very large arbitrary networks.
\end{abstract} 
\section{Introduction}
The information carried by wide area networks is, in general, very important. Yet these networks regularly undergo failures. Detailed statistics about the network failures can be found in \cite{diversitysubMs}. This paper focuses on recovery from single link failures since they consist of 70\% of all network failures. To minimize the cost of such failures, various restoration and protection techniques are developed. The two main metrics in the design of these techniques are restoration speed and capacity efficiency. Capacity efficiency is measured by the total required capacity, in terms of fiber miles, and restoration speed is measured by the duration between the occurrence of failure and restoration of failed traffic. The goal is to minimize both of these metrics and every technique offers a different tradeoff.

In some recovery techniques, spare resources are shared among different traffic failure scenarios and different connection demands, whereas in others, spare resources are dedicated to connection demands. Dedicated protection techniques are able to offer near-hitless recovery since they do not require the signaling and rerouting of the failed traffic. 1+1 Automatic Protection Switching (APS) is a dedicated protection technique where two link-disjoint paths for each connection demand are employed to transmit the same data to the destination node. In the case of a link failure over the primary path, the destination node switches to the protection path and restores the traffic nearly instantaneously. However, 1+1 APS requires more than 100\% capacity which makes it capacity inefficient. The fact that 1+1 APS is currently employed in today's networks \cite{ibbarla} indicates the need for nearly instantaneous link failure recovery despite its low capacity efficiency.

The capacity efficiency of dedicated protection schemes such as 1+1 APS can be improved if the dedicated paths are shared. This can be achieved by employing coding, in particular, erasure coding \cite{aigm}, \cite{aigm2}. The technique introduced in \cite{aigm}, \cite{aigm2}, called diversity coding, has two advantages. First, unlike 1+1 APS, it is capacity efficient. Second, unlike rerouting-based restoration schemes, the recovery is nearly instantaneous. References \cite{aigm}, \cite{aigm2} predate network coding, usually considered to be introduced in \cite{acly}.

In \cite{diversity}, diversity coding is implemented over arbitrary networks using a heuristic algorithm. In \cite{diversitysubMs}, optimal algorithms for the diversity coding technique are developed. Diversity coding performs near-hitless recovery while offering competitive capacity efficiency. In \cite{CPP}, a solution of Shared Path Protection (SPP) \cite{rama} is converted to a coding-based solution named Coded Path Protection (CPP). Sharing of the spare resources is replaced with the employement of these resources to code different paths. This conversion increases the restoration speed and the transmission integrity, and decreases error signaling complexity. The bidirectional nature of CPP allows encoding and decoding inside the network for unicast demands.

In \cite{Kamal2} and \cite{Kamal3}, network coding-based protection schemes called 1+N protection are proposed in which coding operations are carried out over trees and trails, respectively. The idea is similar to that of diversity coding except the protection is bidirectional. In \cite{Hover}, the cost efficiencies of a network coding-based recovery technique and a simpler version of diversity coding technique are evaluated.

All of the above mentioned techniques implement systematic coding where primary paths are exempt from coding operations. Also, in these techniques, coding operations are bound to specific topologies. In addition, they require strict link-disjointness between each primary path and the protection paths. Even though these assumptions make those techniques easier to implement, they have restricted capacity efficiencies. In \cite{ibbarla}, an argument that 1+N coding requires high nodal degree, which reduces its efficiency on sparse topologies was made.

In \cite{diversity_netcod}, the primary paths are incorporated into coding operations using a heuristic algorithm for static provisioning. The decodability of the coding structures is preserved by randomly adding the connection demands to the existing coding groups one by one. A coding group is a set of connection demands that are coded and protected together. Coding primary paths increases capacity efficiency over conventional diversity coding, as in \cite{diversity_netcod}.
Nonsystematic coding is implemented in wireless mesh networks for single link failure recovery in \cite{kamal_many}. In \cite{rouy}, a general network-coding based approach is presented which employs nonsystematic coding and does not explicitly require link-disjointness between primary paths and protection paths. However, this approach is restricted to specific topologies. In addition, it can protect at most two connection demands simultaneously. In \cite{rouyITA}, the proposed technique lifts the restriction over the number of protected connection demands for bidirected networks.
In general, the coding-based recovery techniques in the literature, such as \cite{Kamal2}, \cite{rouy}, \cite{rouyITA}, offer promises in terms of capacity efficiency and restoration time. However, they cannot be optimally implemented on real networks due to their high design complexity limitations. The test networks and traffic matrices in those papers are much smaller than the real networks, such as the long-distance networks of the U.S. and France, to be discussed in sequel. In \cite{diversity_glob}, a novel two step approach is presented to cope with high design complexity on realistic networks. The first step of this algorithm is the pre-processing phase in which all candidate coding groups are calculated and enumerated. In the second step, some of those candidate coding groups are selected and placed on the networks to meet the traffic demand. It manages to overcome the complexity incurred by the size of the traffic matrix. However, the number of coding groups is exponentially dependent on the network size and the nodal degree of the destination node.

This paper offers two novel contributions to the field of diversity coding-based (or network coding-based) link failure recovery. First, we introduce an optimal, simple, and modular design algorithm that provisions the static traffic in very large arbitrary networks. The design algorithm uses column generation technique which does not require explicit enumeration of the coding groups. It starts the problem with a small set of coding groups and creates new coding groups when they are needed. The underlying coding structure of this algorithm is arbitrary as long as the destination nodes of the connections are the same, which offers a solution for different techniques under the same framework. Second, we improve the coding structure of simple diversity coding by offering a coherent coding structure using an Integer Linear Programming (ILP) formulation. In a coherent diversity coding structure, we implement a more relaxed link-disjointness criterion between the paths in a coding group. This enables to form coding groups with higher flexibility and bigger size. The decodability is preserved while the high nodal degree requirement is mitigated. Coherent diversity coding incorporates also nonsystematic coding.

The performance of the new proposed coding technique and the column generation-based design algorithm are investigated compared to conventional (systematic or nonsystematic) diversity coding and {\em p}-cycle protection \cite{GroverBook}. The simplicity of the new design algorithm is also tested based on a set of simulations over relatively large the long-distance networks of the U.S. and France.

\section{Column Generation Method}
The column generation method is an effective technique to solve relatively large linear programming (LP) formulations without explicitly enumerating all possible variables. In some problems, only a small subset of the variables are nonzero in the final solution. In those problems, column generation starts with a small set of variables and creates new and useful variables (columns) which will be likely employed in the final solution. In general, column generation dramatically decreases the time and space complexity depending on the nature of the problem. In the network-coding based link failure problem, column generation technique results in significant time and memory savings, and therefore it enables the optimal implementation of efficient network coding-based techniques over very large realistic networks.

Column generation has been used for different LP problems, including the well-known cutting stock problem \cite{ColumnBook}. The problem is to satisfy different widths of paper demand by cutting fixed width rolls in different patterns. The goal is to use a minimum number of rolls. The problem starts with a small set of basic cutting patterns. The useful cutting patterns are generated one-by-one. We observed that the diversity coding-based link failure recovery problem is very similar to the cutting stock problem. Diversity coding over arbitrary networks can be implemented like the cutting stock problem as long as the cutting patterns are replaced by coding groups and the demands for different widths of paper are replaced with the traffic demands of a single destination node. The only difference is the fact that coding groups can have different costs, whereas in the cutting stock problem, each cutting pattern is cut from rolls with the same total width. Other advanced methods developed for the cutting stock problem can also be applied to the diversity coding implementation.

The column generation technique is also applied to the {\em p}-cycle protection \cite{pcycle_col} and SPP \cite{SPP_col} techniques resulting in significant time and memory savings. It is a better fit to diversity coding technique than {\em p}-cycle protection and SPP since, in diversity coding, there is a single subproblem that generates coding groups. However, in {\em p}-cycle protection, there is a subproblem for both generating {\em p}-cycles and generating candidate paths for each connection demand. Likewise, in SPP, there is a different subproblem for generating candidate path pairs for each connection demand.

\ifCLASSOPTIONonecolumn
\begin{figure}[t!]
\centering
\includegraphics[bb = 52 60 693 445, width=120mm]{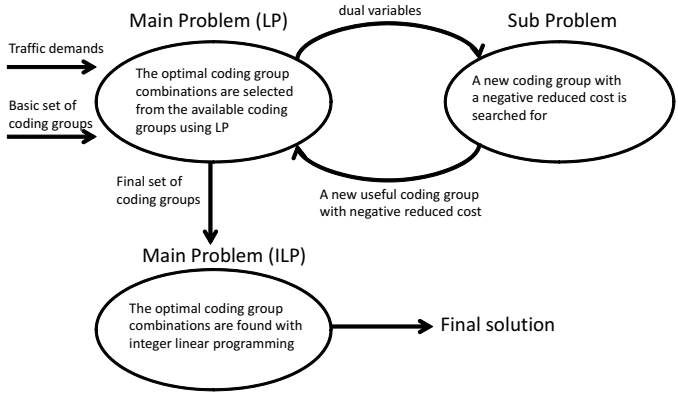}
\caption{Steps in the column generation method.}
\label{fig:CPP design}
\end{figure}
\else
\begin{figure}[!t]
\centering
\includegraphics[bb = 52 60 693 445,width=90mm,clip=true]{columngeneration1.eps}
\caption{Steps in the column generation method.}
\label{fig:CPP design}
\end{figure}
\fi

The column generation method for diversity coding is visualized in Fig.~\ref{fig:CPP design}. There are two main components of this method. The main problem, which is also called the \textit{Coding Groups Placement Problem}, inputs the traffic demands and a set of basic coding group. The main problem in this step is an LP formulation that finds the optimal coding group combinations to meet the traffic demands. After the first run, it passes the dual variables of the solution to the subproblem. The subproblem, which is also called the \textit{Coding Group Generation Problem}, attempts to find a new useful coding group. A new useful coding group has a negative reduced cost given the dual variables of the main problem. The new useful coding group is input to the main problem iteratively. In the next round, optimal coding group combinations are found given the expanded coding group set. The dual variables of this run are input to the subproblem as before. This iterative operation is carried out until the subproblem cannot find any new coding group with a negative reduced cost. The main problem is then solved one last time as an ILP. The gap between ILP and LP solutions of the main problem is generally very small, as will be discussed in Section \ref{sec:Results}.
\subsection{Example 1}
As an example, assume there are 2 connection demands from $S_1$ and $S_2$ to $D$. Each has a unit traffic demand. The cost of coding groups that protect only $S_1-D$ and only $S_2-D$ are 10 and 7, respectively. The coding groups combination problem employs one of each coding group to satisfy the traffic demands at a total cost of 17. The dual variables of this solution are input to the subproblem. The subproblem returns a new coding group consisting of both $S_1-D$ and $S_2-D$ at a total cost of 15. The main problem is run one more time and decreases the total cost from 17 to 15 since it employs only the new coding group created by the subproblem. The optimal result is achieved since the subproblem cannot create any more coding groups with negative reduced costs.

\section{ILP Formulations}
In this section, we present the algorithms that realize the main problem and the subproblem. The main problem finds the optimal combination of coding groups out of a given set and places them on the network to meet the traffic demands. Throughout the iterative process, the main problem is realized with an LP formulation, whereas in the last step, the formulation is converted to an ILP since in the final solution coding groups must be replaced in integer numbers. On the other hand, the realization of the subproblem is not unique. The coding group generation algorithm depends on the adopted coding structure. In addition, the way new coding groups are generated can be realized by heuristic techniques. In this section, we present three different coding group generation algorithms using mixed integer programming (MIP) or ILP formulations.

\subsection{Main Problem (Coding Groups Placement Problem)}
\label{sec:Main}
An LP formulation is developed to implement the coding groups placement algorithm, which serves the main problem of the column generation method. The goal is to place a coding group with minimum total cost while meeting the traffic demands. The input parameters of the LP are

\begin{itemize}
\item $CG$    : The set of coding groups, this set is expanded at each iteration,
\item $V$     : The set of nodes,
\item $t_f$   : The traffic demand from source node $f$ to destination node $d$,
\item $cost_i$ : The cost of coding group $i$,
\item $CG_{i,f}$ : The number of connections originating from node $f$ in coding group $i$.
\end{itemize}
The variables related to the coding groups placement problem can be either continuous or integer
\begin{itemize}
\item $n(i)$   : Keeps the number instances of coding group $i$ placed on the network, normally continuous.
\end{itemize}
The variables $n(i)$ are converted to integer variables at the final step of column generation.

The objective function is
\begin{equation}
\min \sum_{i\in CG} cost_i\times n(i).
\end{equation}
The following inequalities ensure a sufficient number of coding groups are placed to protect all of the
traffic demands
\begin{equation}
\sum_{i\in CG} CG_{i,f} \times n(i) \geq t_f \hspace{5mm} \forall f \in V, f\neq d,
\end{equation}
The diagram of the column generation method in terms of the parameters and variables of the LP formulations is shown in Fig.~\ref{fig:column2}, where $\pi_f$ are the dual variables of each constraint.
\ifCLASSOPTIONonecolumn
\begin{figure}[t!]
\centering
\includegraphics[bb = 52 60 693 445, width=120mm]{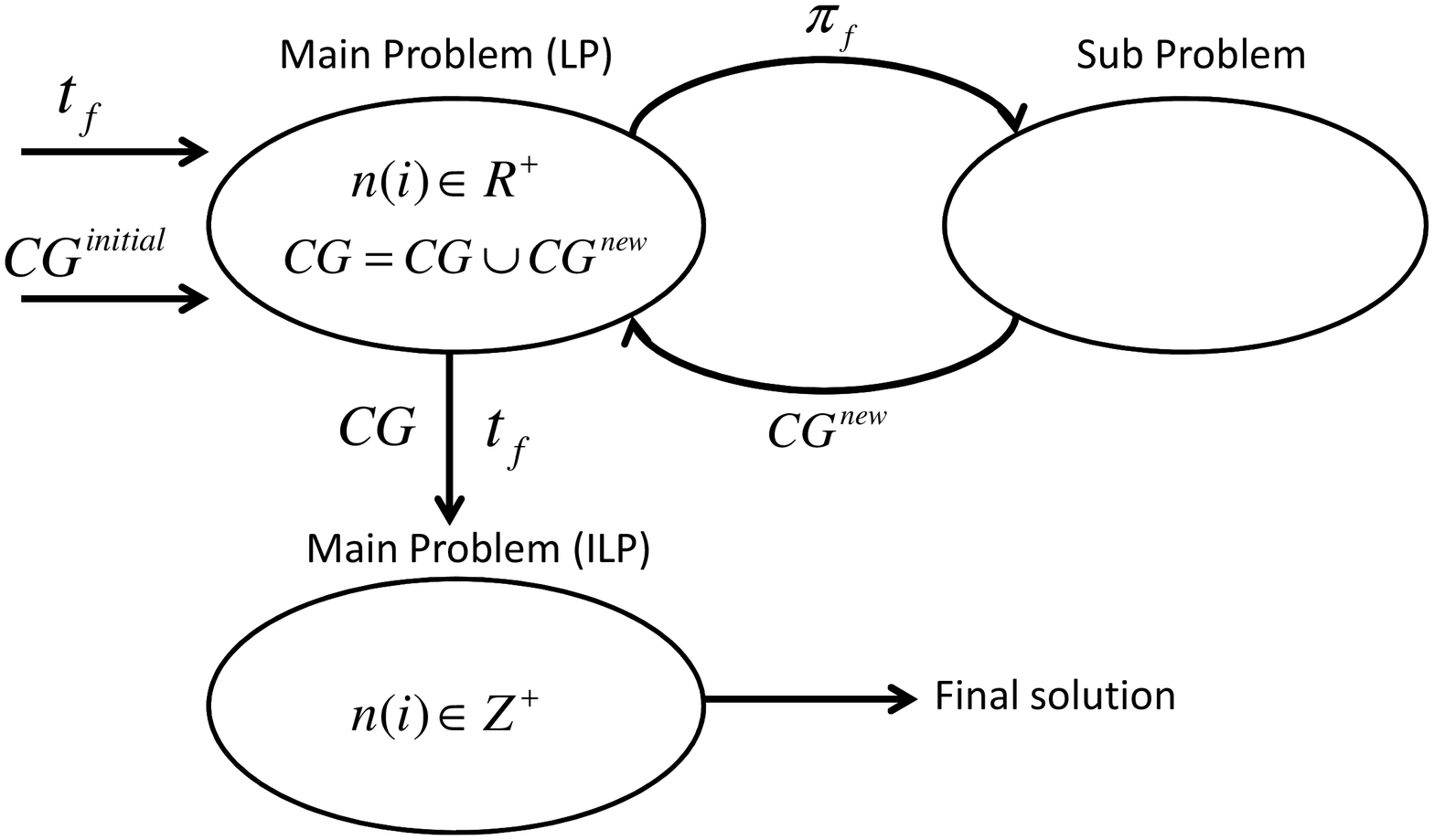}
\caption{Steps in the column generation method in terms of LP and ILP variables.}
\label{fig:column2}
\end{figure}
\else
\begin{figure}[!t]
\centering
\includegraphics[bb = 52 60 693 445,width=90mm,clip=true]{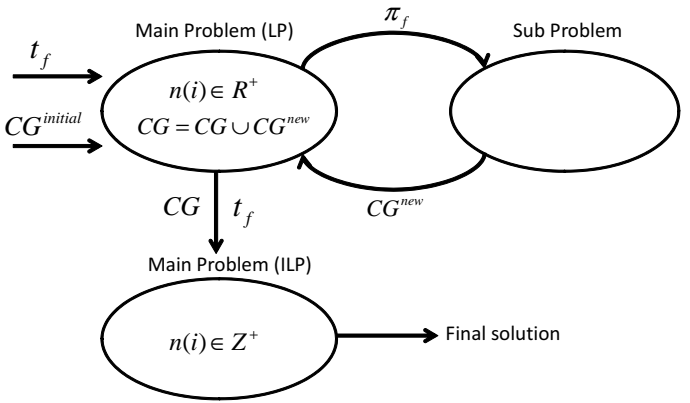}
\caption{Steps in the column generation method in terms of LP and ILP variables.}
\label{fig:column2}
\end{figure}
\fi
The traffic demand parameters $t_f$ and an initial basic coding group set $CG^{initial}$ are input to the main problem. After the first run, the main problem inputs the resulting dual values of the constraints to the subproblem. The subproblem returns a new coding group with negative reduced cost, if available. The iterative process terminates when the subproblem cannot produce any more new coding groups with reduced cost. Then the variables $n(i)$ are converted to integer variables and ILP is run at the last step to get the final solution.

\subsection{Subproblem (Coding Group Generation Problem)}
The objective of the subproblem is to find a new coding group in each iteration that will be useful in the main problem. The subproblem inputs the dual variables of the main problem and returns a new coding group. A new coding group can be selected among many which have negative reduced costs. In this paper, we opt to search for a new coding group with the minimum negative reduced cost until there is at least one. We present three different coding group generation algorithms, each implementing a different version of diversity coding. These versions have a tradeoff of simplicity versus capacity efficiency. In the following subsections, they are presented in increasing order of capacity efficiency and design complexity.
\subsubsection{Systematic Diversity Coding}
In this algorithm, we adopt systematic diversity coding where only protection paths are encoded. The core algorithm is adopted from the diversity coding tree algorithm in \cite{diversity_journal}. In a coding group, there is a primary tree serving as the union of the primary paths of the protected connections. There is also a link-disjoint protection tree whose branches originate from the source nodes of the protected connections. Those branches merge when they come together until they achieve the destination node. An example is taken from \cite{diversity_journal} and is shown in Fig.~\ref{fig:dtree1}. There are three connection demands originating from $S_1$, $S_2$, and $S_3$ going to node $D$. The solid black lines represent the primary tree whereas dashed lines represent the protection tree.
\ifCLASSOPTIONonecolumn
\begin{figure}[h!]
\centering
\subfigure[]{
\includegraphics[width=75mm]{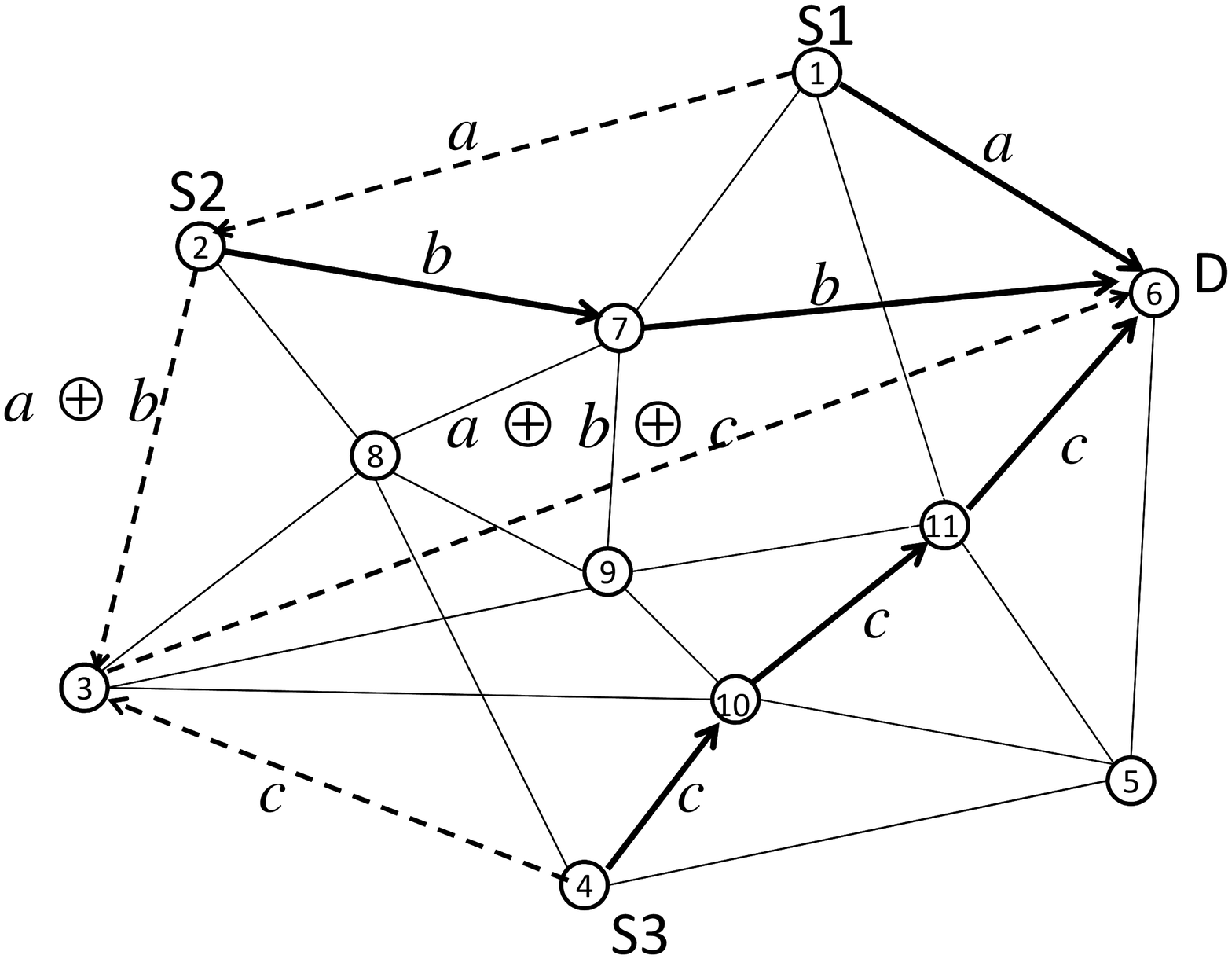}
\label{fig:dtree1}
}
\subfigure[]{
\includegraphics[width=75mm]{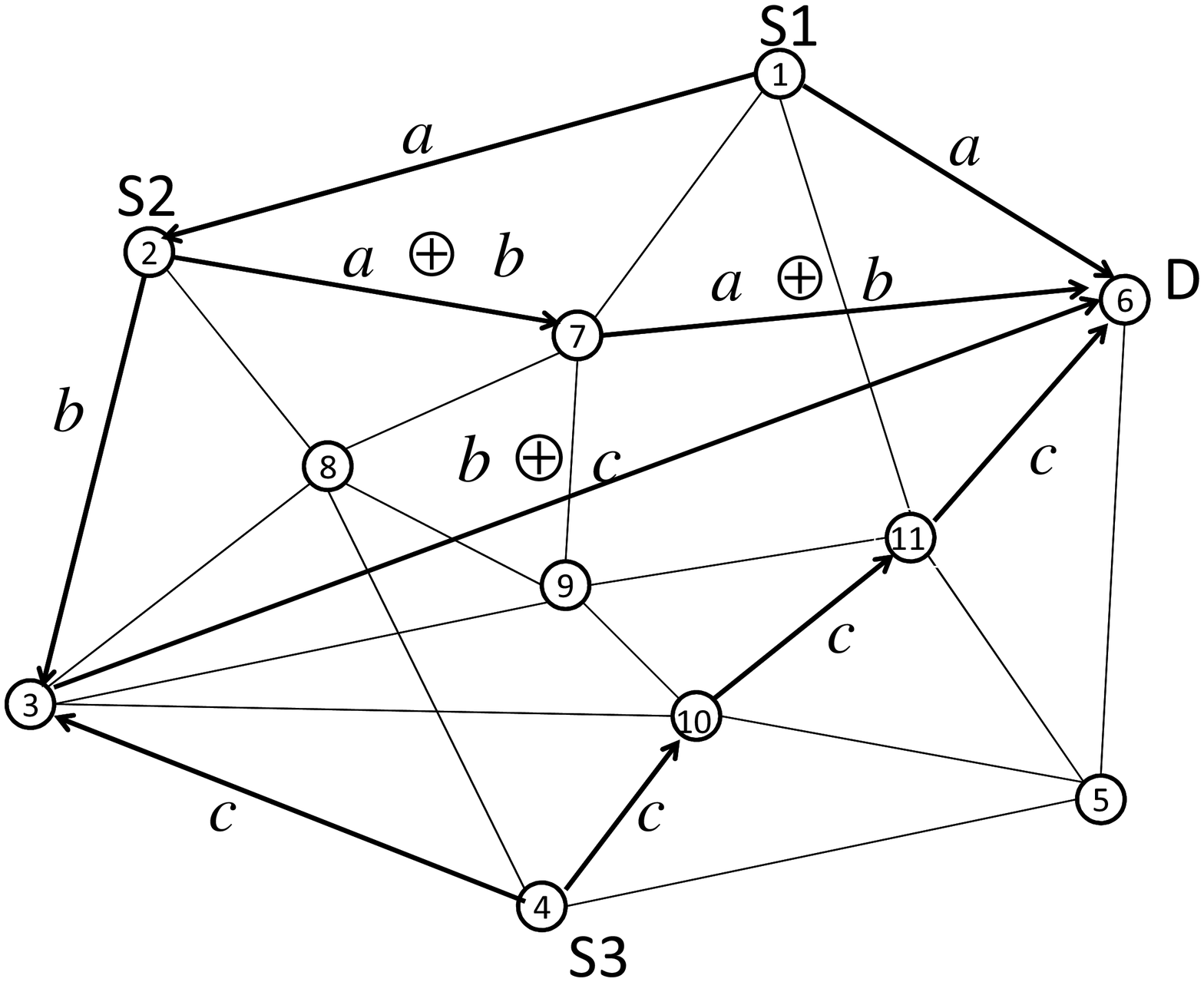}
\label{fig:dtree2}
}
\caption[Optional caption for list of figures]{\subref{fig:dtree1} An example of the systematic diversity coding tree structure. There are three link-disjoint primary paths spanned by the primary tree and there is a link-disjoint protection tree, \subref{fig:dtree2} An example of nonsystematic diversity coding structure for the same set of connections.}
\label{fig:dtree}
\vspace{-5mm}
\end{figure}
\else
\begin{figure}[!t]
\centering
\subfigure[]{
\includegraphics[width=75mm]{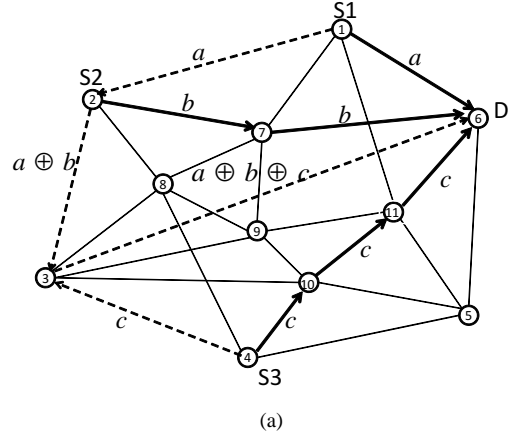}
\label{fig:dtree1}
}
\subfigure[]{
\includegraphics[width=75mm]{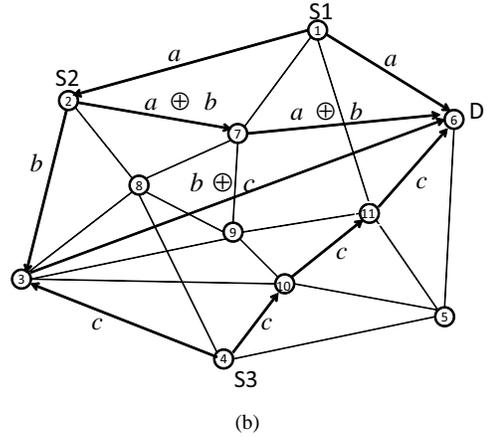}
\label{fig:dtree2}
}
\caption[Optional caption for list of figures]{\subref{fig:dtree1} An example of the systematic diversity coding tree structure. There are three link-disjoint primary paths spanned by the primary tree and there is a link-disjoint protection tree, \subref{fig:dtree2} An example of nonsystematic diversity coding structure for the same set of connections.}
\label{fig:dtree}
\end{figure}
\fi

The input parameters required in the MIP formulation of the coding group generation algorithm based on systematic diversity coding are
\begin{itemize}
\item $G(V,E)$   : Network graph,
\item $S$        : The set of spans in the network, a span consists of two links in the opposite directions,
\item $a_e$      : Cost associated with link $e$,
\item $\Gamma_i(f)$ : The set of incoming links of each node $f$,
\item $\Gamma_o(f)$ : The set of outgoing links of each node $f$,
\item $d$        : The common destination node,
\item $ND$     : The nodal degree of the destination node $d$,
\item $\alpha$   : A constant employed in the algorithm where $\frac{1}{|V|}\geq \alpha \geq 0$,
\item $\beta$    : A constant employed in the algorithm, $\beta \geq 2\times \max(|V|, \max_i(ND_i))$,
\item $\pi_f$    : The values of the dual variables of the main problem.
\end{itemize}
The set of variables of this MIP formulation are
\begin{itemize}
\item $CG_f^{new}$  : Integer variable, equals to the number of connections originating from node $f$ in the new coding group,
\item $d_e \in \{0-1\}$ : Integer variable, equals $1$ iff the primary tree of the new coding group passes through link $e$,
\item $c_e \in \{0-1\}$ : Integer variable, equals $1$ iff the protection tree of the new coding group passes through link $e$,
\item $p_f$   : A continuous variable between $0$ and $1$. It keeps the ``voltage'' value of node $f$ in the protection tree of the new coding group.
\item $g_f$   : Same description as $p_f$ except it is used for the primary tree of the new coding group.
\end{itemize}

The objective function minimizes the reduced cost of a new coding group
\begin{equation}
\min \sum_{e\in E}(d_e+c_e)\times a_e - \sum_{f\in V} CG_f^{new}\times \pi_f.
\end{equation}
If the value objective function comes out to be negative then a new coding group is found and
input to the main problem.
\begin{equation}
\sum_{f\in V} CG_f^{new} \leq ND-1 \hspace{5mm}\forall i ,
\label{ineq:bound}
\end{equation}
\begin{equation}
\sum_{e\in\Gamma_o(f)} d_e = CG_f^{new}+\sum_{e\in\Gamma_i(f)}d_e\hspace{3mm} \forall f \in V, f\neq d, \hspace{5mm}
\label{eq:pri3}
\end{equation}
\begin{equation}
\sum_{e\in \Gamma_i(d)} d_e = \sum_{f\in V} CG_f^{new},\hspace{5mm}
\label{eq:pri1}
\end{equation}

\begin{equation}
\sum_{e\in\Gamma_o(d)} d_e+c_e = 0,\hspace{5mm}
\label{eq:pri2}
\end{equation}

\begin{equation}
\sum_{e\in\Gamma_o(f)} c_e \ge \frac{\displaystyle CG_f^{new}}{\beta} +\frac{\displaystyle\sum_{e\in\Gamma_i(f)} c_e}{\beta}\hspace{5mm}\forall f\in V, f\neq d.
\label{eq:pro3}
\end{equation}

\begin{equation}
\sum_{e\in\Gamma_i(d)} c_e \ge \frac{\displaystyle\sum_{f\in V}CG_f^{new}}{\beta},\hspace{5mm}
\label{eq:pro1}
\end{equation}

\begin{equation}
d_{e1}+d_{e2}+c_{e1}+c_{e2}\le 1\hspace{5mm}\forall e1,e2 \in g, \forall g\in S,
\label{eq:dis1}
\end{equation}

\begin{equation}
g_f-g_f\ge \alpha \cdot d_e - (1-d_e)\hspace{5mm}\forall e \in E,
\label{eq:vol1}
\end{equation}

\begin{equation}
p_f-p_f\ge \alpha \cdot c_e - (1-c_e)\hspace{5mm}\forall e \in E.
\label{eq:vol2}
\end{equation}
Inequality (\ref{ineq:bound}) ensures that the size of the new coding group does not exceed ND-1. Equation (\ref{eq:pri3}) carries out the origination and continuation of the primary tree, whereas equation (\ref{eq:pri1}) and equation (\ref{eq:pri2}) carry out the termination of the primary tree. Inequality (\ref{eq:pro3}) is responsible for the origination and continuation of the protection tree, whereas inequality (\ref{eq:pro1}) and equation (\ref{eq:pri2}) are responsible for the termination of the protection trees. Inequality (\ref{eq:dis1}) makes sure that primary and protection trees are link-disjoint. Inequalities (\ref{eq:vol1}) and (\ref{eq:vol2}) assign voltage values to nodes to prevent getting cyclic structures in primary and protection trees, respectively.

\subsubsection{Nonsystematic Diversity Coding}
\label{sec:Nonsys}
In this section, the coding groups are generated based on a more generic coding structure where both primary
and protection paths can be encoded. We refer to \textit{Lemma 1} from \cite{kamal_many} while building valid nonsystematic diversity coding. This coding structure increases the capacity efficiency of systematic diversity coding with extra design complexity. An example is shown in Fig.~\ref{fig:dtree2}. Different from systematic diversity coding, the primary paths of $S_1-D$ and $S_2-D$ are encoded. The core algorithm to generate new coding groups in the column generation method is an ILP formulation taken from \cite{diversity_journal} with small changes. Reference \cite{diversity_journal} presents how to optimally build nonsystematic diversity coding structures. The algorithm in \cite{diversity_journal} looks for every possible coding scenario by eliminating the invalid cases that can be identified as \textit{coding cycles.}
The ILP formulation of the nonsystematic diversity coding group generation algorithm has a set of binary integer variables taking values from the set $\{0,1\}$
\begin{itemize}
\item $x_e(i)$   : Equals 1 iff the path $i$ passes through link $e$,
\item $n(i,s)$   : Equals 1 iff path $i$ is in subgroup $s$,
\item $m(i,j)$   : Equals 1 iff path $i$ and path $j$ are in the same subgroup so are coded together,
\item $r(i,f)$   : Equals 1 iff path $i$ and connection demand $f$ are indirectly related,
\item $t_e(s)$   : Equals 1 iff one of the paths in subgroup $s$ traverses over link $e$,
\item $\sigma_{f,i}$ : Equals 1 iff node $f$ is the source node of demand $i$.
\end{itemize}
The objective function is
\begin{equation}
\min \sum_{e\in E}\sum_{s=1}^{2N} t_e(s)\times a_e - \sum_{f\in V} CG_f^{new}\times \pi_f.
\end{equation}

\begin{equation}
\sum_{f\in V} \sigma_{f,i}\leq 1, \hspace{5mm} 1\leq i\leq ND-1 :\bmod(i,2)=0, \label{eq:non1}
\end{equation}
\begin{equation}
CG_f^{new} = \sum_{i=1}^{ND-1} \sigma_{f,i} \hspace{5mm} \forall f\in V, f\neq d, \label{eq:non2}
\end{equation}
\begin{equation}
\sum_{f\in V} \sum_{i=1}^{ND-1}, \sigma_{f,i} \leq ND-1 \label{eq:non3}
\end{equation}
\ifCLASSOPTIONonecolumn
\begin{equation}
\sum_{e\in\Gamma_i(f)}x_e(j)-\sum_{e\in\Gamma_o(f)}x_e(j)=\left\{\begin{tabular}{ll}
$-\sigma_{f,i}$&{\rm if\ }$v\neq d,$\\
$\sum \sigma_{f,i}$&{\rm if\ }$v=d,$\end{tabular}
\hspace{3mm}j = 2i,\hspace{1mm} j = 2i-1.
\right.
\label{eq:non6}
\end{equation}
\else
\begin{eqnarray}\nonumber
\sum_{e\in\Gamma_i(f)}x_e(j)-\sum_{e\in\Gamma_o(f)}x_e(j)=\left\{\begin{tabular}{ll}
$-\sigma_{f,i}$&{\rm if\ }$v\neq d,$\\
$\sum \sigma_{f,i}$&{\rm if\ }$v=d,$\end{tabular}\right. \\
j = 2i,\hspace{1mm} j = 2i-1.
\label{eq:non6}
\end{eqnarray}
\fi
\begin{equation}
\sum_{s=1}^{2(ND-1)} n(i,s)=1, \hspace{5mm} 1\leq i\leq 2(ND-1),
\label{eq:non5}
\end{equation}
\ifCLASSOPTIONonecolumn
\begin{equation}
n(i,s) + n(i-1,s) \leq 1, \hspace{5mm} 1\leq i,s\leq 2(ND-1) : \bmod(i,2)=0, \label{eq:non4}
\end{equation}
\else
\begin{eqnarray}\nonumber
n(i,s) + n(i-1,s) \leq 1, \\
1\leq i,s\leq 2(ND-1) : \bmod(i,2)=0, \label{eq:non4}
\end{eqnarray}
\fi
\begin{equation}
t_e(s) \geq x_e(i) + n(i,s) - 1 \hspace{3mm} \forall e,i,s \label{eq:non7}
\end{equation}
\ifCLASSOPTIONonecolumn
\begin{equation}
t_e(s_1) + t_e(s_2) + t_k(s_1) + t_k(s_2) \leq 1 \hspace{5mm} \forall e,k \in g, \forall g\in S, \forall s_1,s_2 \label{eq:non8}
\end{equation}
\else
\begin{eqnarray}\nonumber
t_e(s_1) + t_e(s_2) + t_k(s_1) + t_k(s_2) \leq 1\\ 
\hspace{5mm} \forall e,k \in g, \forall g\in S, \forall s_1,s_2  \label{eq:non8}
\end{eqnarray}
\fi
\begin{equation}
m(i,j) \geq n(i,s) + n(j,s)-1 \hspace{5mm} \forall i\neq j,s.  \label{eq:non9}
\end{equation}
\ifCLASSOPTIONonecolumn
\begin{eqnarray}
r(i,f) \geq m(i,j)+m(j^*,2f)+m(j^*,2f-1)-m(i,2f) -m(i,2f-1)-1 \hspace{5mm} \forall i,j,f,:i\neq j \label{eq:non10}
\end{eqnarray}
\else
\begin{eqnarray}\nonumber
r(i,f) \geq m(i,j)+m(j^*,2f)+m(j^*,2f-1)\\
-m(i,2f) -m(i,2f-1)-1 \hspace{5mm} \forall i,j,f,:i\neq j \label{eq:non10}
\end{eqnarray}
\fi
such that $j^*=j-1$ if $\bmod{(j,2)}=0$ and $j^*=j+1$ otherwise.
\ifCLASSOPTIONonecolumn
\begin{eqnarray}\nonumber
r(i,f) \geq r(i,g)+m(2g,2f)+m(2g,2f-1)+m(2g-1,2f) +m(2g-1,2f-1)-1 \hspace{2mm} \\
\forall i,f\neq g:i\neq 2f, i\neq 2f-1,i\neq 2g, i\neq 2g-1. \label{eq:non11}
\end{eqnarray}
\else
\begin{eqnarray}\nonumber
r(i,f) \geq r(i,g)+m(2g,2f)+m(2g,2f-1)\\
\nonumber +m(2g-1,2f) +m(2g-1,2f-1)-1 \hspace{2mm} \forall i,f\neq g: \\
i\neq 2f, i\neq 2f-1,i\neq 2g, i\neq 2g-1. \label{eq:non11}
\end{eqnarray}
\fi
\ifCLASSOPTIONonecolumn
\begin{eqnarray}\nonumber
r(2f,g) + r(2f-1,g)+m(2f,2g)+m(2f-1,2g)+m(2f,2g-1) +m(2f-1,2g-1)\leq 1 \\
\hspace{2mm} \forall g,f : g\neq f, \label{eq:non12}
\end{eqnarray}
\else
\begin{eqnarray}\nonumber
r(2f,g) + r(2f-1,g)+m(2f,2g)+m(2f-1,2g) \\
+m(2f,2g-1) +m(2f-1,2g-1)\leq 1 \hspace{2mm} \forall g,f : g\neq f, \label{eq:non12}
\end{eqnarray}
\fi
Inequality (\ref{eq:non1}) ensures that each demand has at most one source node. Some connection demands may be empty. Equation (\ref{eq:non2}) calculates the number of connection demands originating from each node at the new coding group. Inequality (\ref{eq:non3}) bounds the total number of connection demands in the new coding group by the nodal degree of the destination node minus 1. Equation (\ref{eq:non6}) carries out the origination, continuation and termination of the paths of each connection demand. Each connection demand has two paths in a coding group. Equation (\ref{eq:non5}) ensures that each connection demand is a part of coding subgroup. Inequality (\ref{eq:non4}) ensures that paths belonging to the same connection cannot be a part of the same subgroup. Inequality (\ref{eq:non7}) compiles the topologies of the subgroups by combining the paths of the demands in that subgroup. Inequality (\ref{eq:non8}) satisfies the link-disjointness criterion between the topologies of different subgroups. Inequality (\ref{eq:non9}) says that if two paths are in the same subgroup then they are assumed to be coded together.
In inequality (\ref{eq:non10}), path $i$ becomes indirectly related to demand $f$ if there exists a path $j$ that is coded with both path $i$ and one of the paths carrying demand $f$. Moreover, path $i$ must not be coded with either paths of demand $f$. In inequality (\ref{eq:non11}), path $i$ becomes indirectly related to demand $f$ if there exists a demand $g$ that is indirectly related to path $i$, and one of the paths of demand $g$ must be coded together with one of the paths of demand $f$. Inequality (\ref{eq:non12}) ensures that two different connection demands can either be indirectly related or one of their paths are encoded together. Otherwise, a coding cycle occurs which is a violation of the validity of the coding structure.
\subsubsection{Coherent Diversity Coding}
\label{sec:coherent}
In this section, we introduce a novel coding structure that can mitigate the limiting link-disjointness criterion to the optimal extent. It is called \textit{Coherent Diversity Coding}. This coding structure is optimal under the conditions listed as
\begin{itemize}
\item There is a single destination node,
\item There are two link-disjoint paths for each connection demand,
\item The coding operations are within $GF(2)$.
\end{itemize}
It enables one to achieve more capacity efficient results than typical diversity coding. Typical diversity coding, systematic or nonsystematic, requires two paths to be either coded or to be link-disjoint. For example, that prevents applying typical diversity coding at the destination nodes with a nodal degree of 2, even though the rest of the network is highly connected. There is a space for improvement in the capacity efficiency of coding groups by relaxing the link-disjointness criterion between different paths. Fig.~\ref{fig:disjointExample} is taken from \cite{rouy} and shows how the strict link-disjointness criterion for two connections can be relaxed in order to save capacity. The connection demands are from node $s$ to node $t$, carrying signals $p_1$ and $p_2$, respectively. There is no available nontrivial solution for diversity coding on this topology since there are only $N$, which is $2$ in this case, number of link-disjoint paths, less than the required $N+1$ ($N+1=3$), from source to destination. Therefore, in Fig.~\ref{fig:disjoint1}, the solution of diversity coding is identical to that of 1+1 APS. The low nodal degree of the source node is a bottleneck for diversity coding. On the other hand, the network-coding based technique proposed by \cite{rouy} shows that these two data signals can be coded to save capacity in Fig.~\ref{fig:disjoint2}. However, the technique  in \cite{rouy} is nontractable for more than two connection demands and lacks an efficient capacity placement algorithm.

Therefore, we developed the optimal link-disjointness criteria between paths in the same coding group that can mitigate the effects of low nodal degree in the network. The coherent diversity coding enables paths sharing the same link,  even if they are not coded together, up to the extent that decodability is preserved. Therefore, it is both optimal and feasible. Under the optimality conditions stated above, the necessary and sufficient conditions of decodability are to ensure that at least one copy of each signal is alive and any subset of $k$ signals resides in at least $k$ subgroups after any single link failure. The resulting coding structure will be decodable according to \textit{Lemma 1} in \cite{kamal_many}. Therefore, we build the coding structure of coherent diversity coding such that after any single link failure, there will be at least one copy of each signal and any subset of $k$ signals reside in at least $k$ subgroups. The terms of coherent and noncoherent paths are coined to keep the track of link-disjointness relationship between paths. If two paths are coherent to each other, then they can fail simultaneously, therefore they can share the same links. Otherwise, their simultaneous failure will impair the decodability as will be shown with an example.
The proposed technique is nearly as simple to implement as diversity coding.

\ifCLASSOPTIONonecolumn
\begin{figure}[h!]
\centering
\subfigure[]{
\includegraphics[width = 60mm]{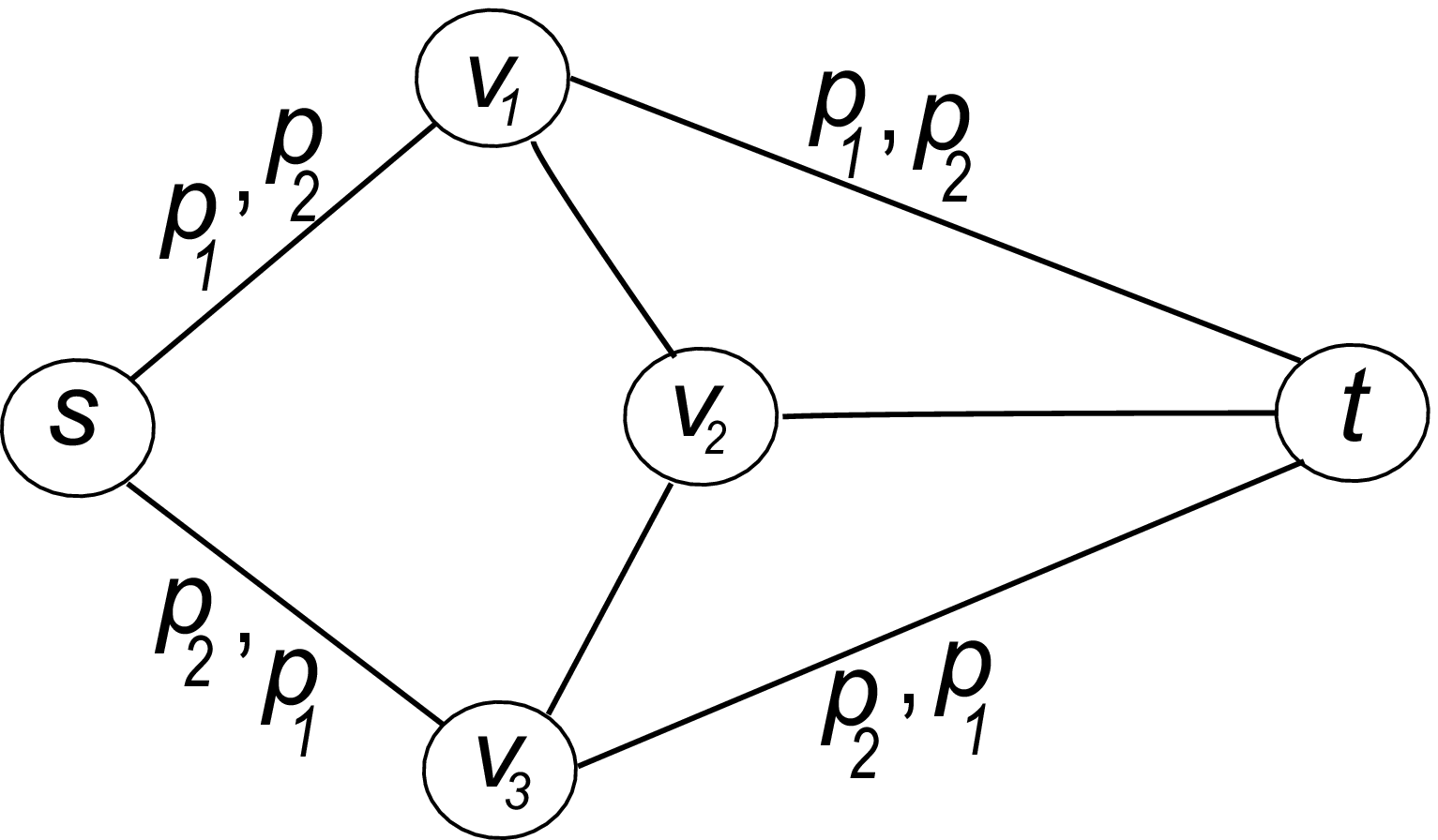}
\label{fig:disjoint1}
}
\subfigure[]{
\includegraphics[width = 60mm]{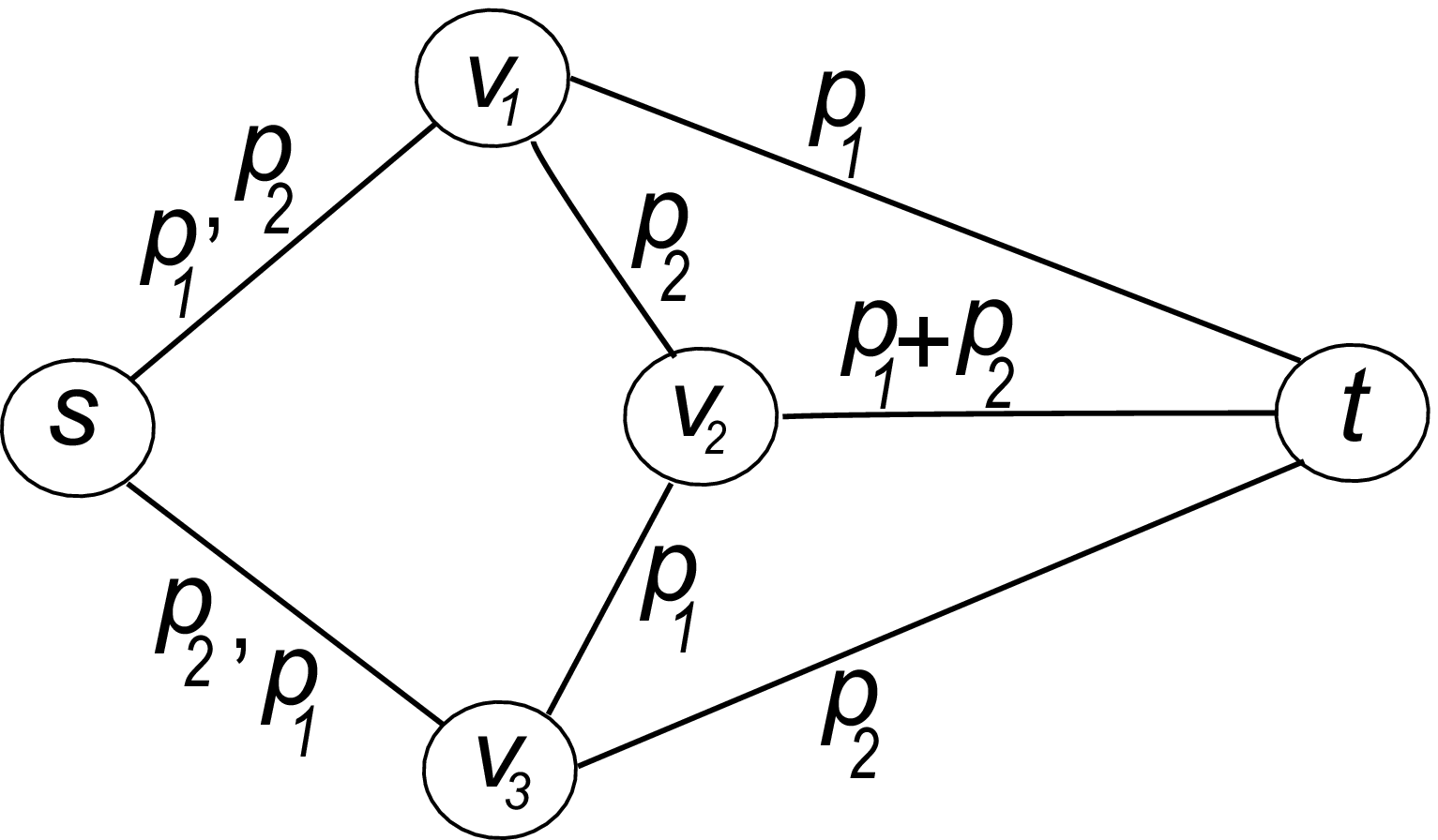}
\label{fig:disjoint2}
}
\caption[Optional caption for list of figures]{Effect of low nodal degree on coding \subref{fig:disjoint1} Diversity coding solution (identical to 1+1 APS), \subref{fig:disjoint2} Network coding-based solution \cite{rouy}.}
\label{fig:disjointExample}
\end{figure}
\else
\begin{figure}[!t]
\centering
\subfigure[]{
\includegraphics[width = 40mm]{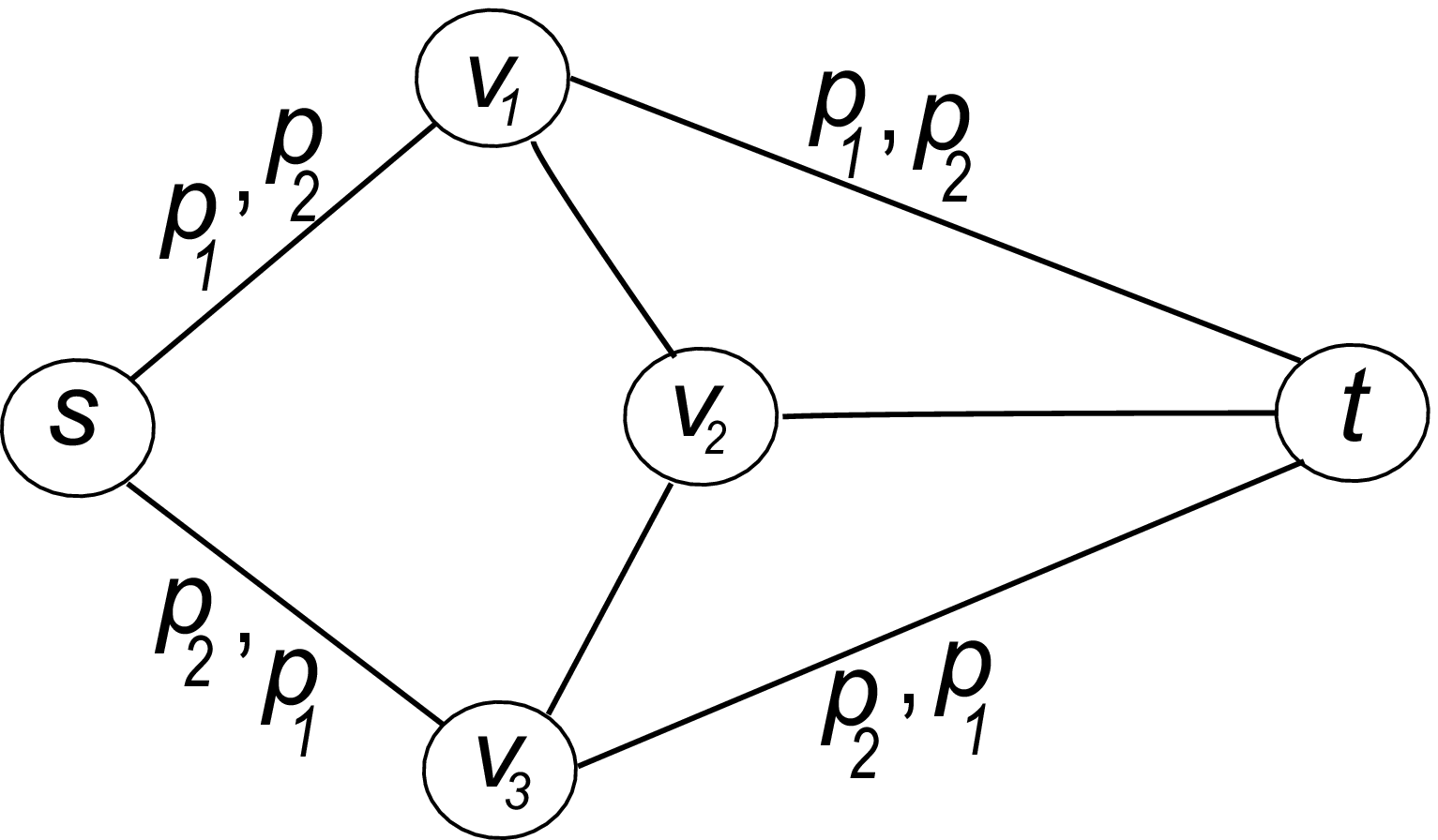}
\label{fig:disjoint1}
}
\subfigure[]{
\includegraphics[width = 40mm]{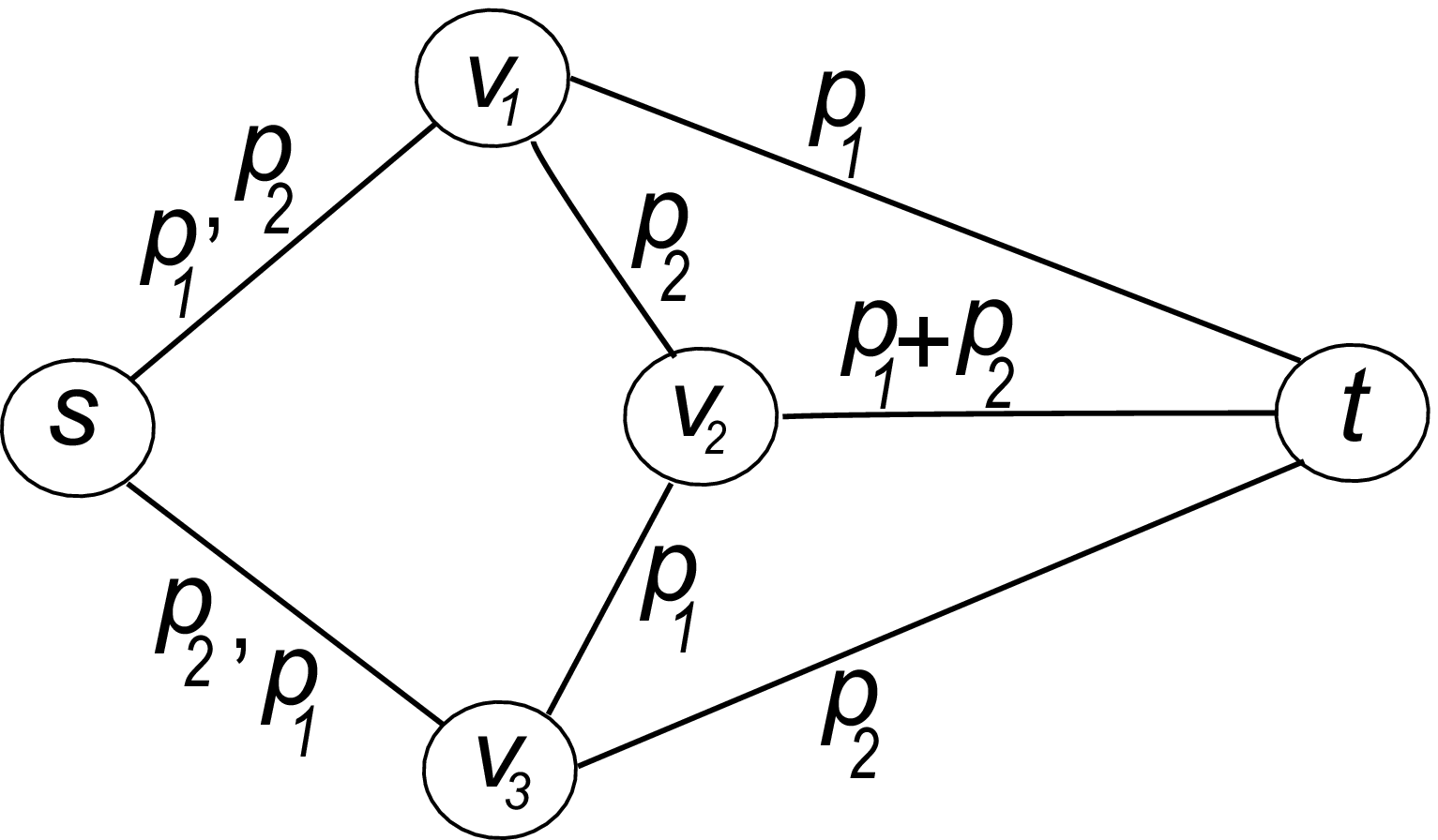}
\label{fig:disjoint2}
}
\caption[Optional caption for list of figures]{Effect of low nodal degree on coding \subref{fig:disjoint1} Diversity coding solution (identical to 1+1 APS), \subref{fig:disjoint2} Network coding-based solution \cite{rouy}.}
\label{fig:disjointExample}
\vspace{-2mm}
\end{figure}
\fi

\ifCLASSOPTIONonecolumn
\begin{figure*}[h!]
\centering
\subfigure[]{
\includegraphics[width = 20mm]{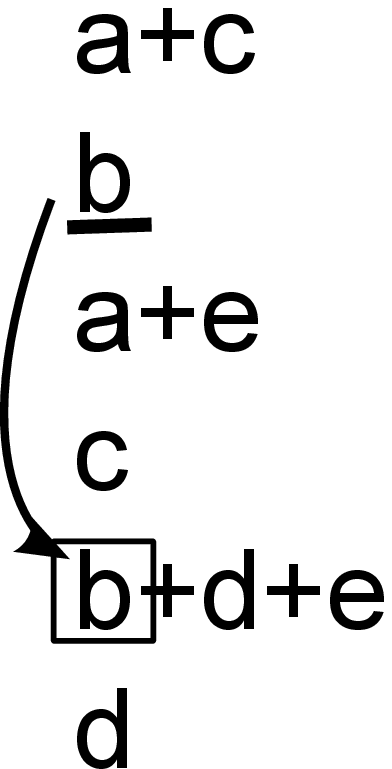}
\label{fig:coh1}
}
\subfigure[]{
\includegraphics[width = 22mm]{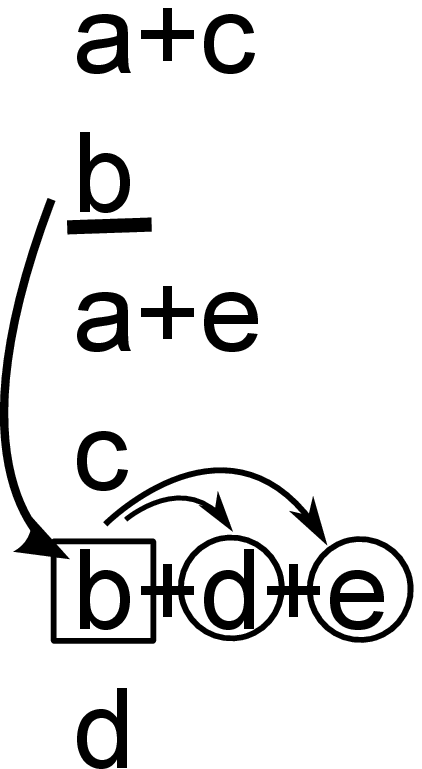}
\label{fig:coh2}
}
\subfigure[]{
\includegraphics[width = 22mm]{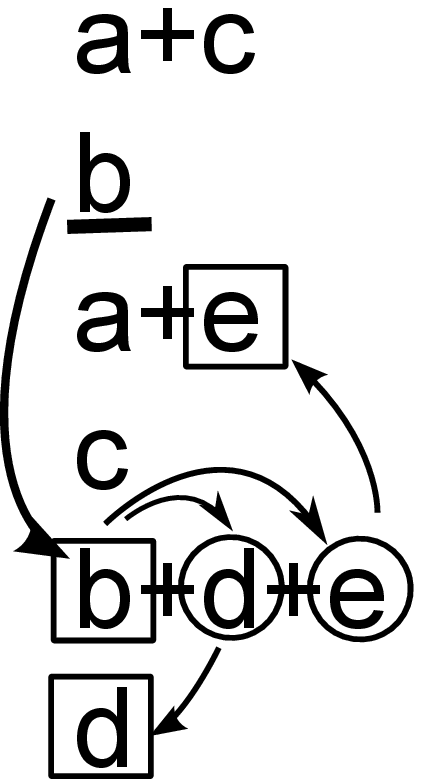}
\label{fig:coh3}
}
\subfigure[]{
\includegraphics[width = 22mm]{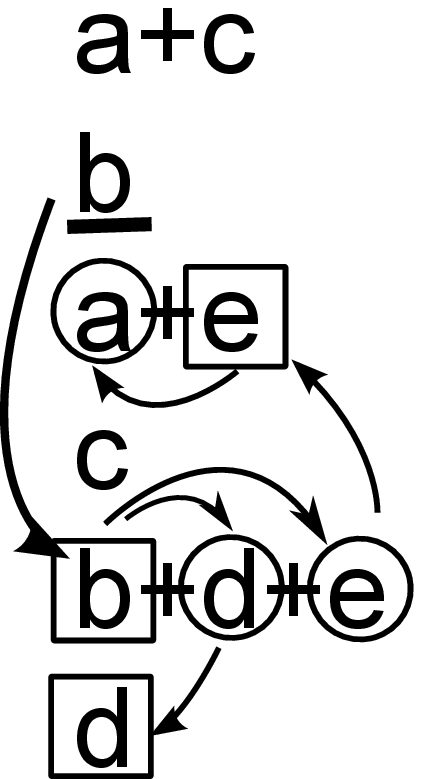}
\label{fig:coh4}
}
\subfigure[]{
\includegraphics[width = 23mm]{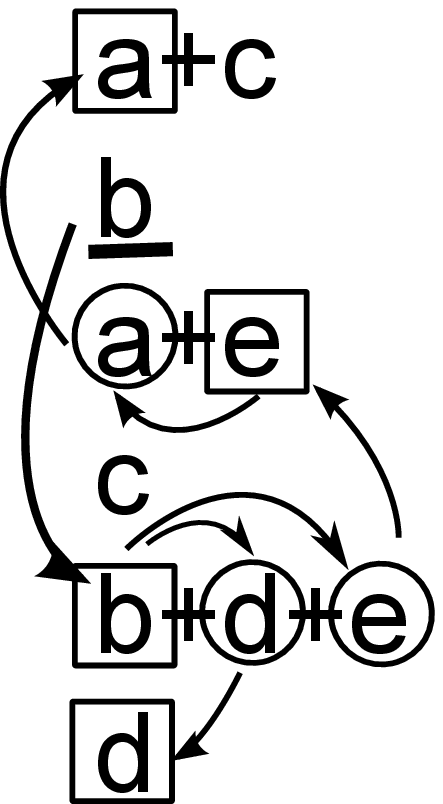}
\label{fig:coh5}
}
\subfigure[]{
\includegraphics[width = 22.7mm]{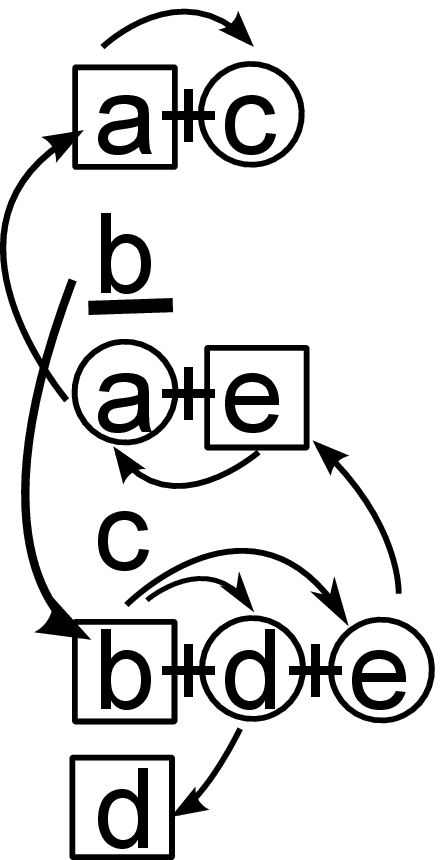}
\label{fig:coh6}
}
\subfigure[]{
\includegraphics[width = 25mm]{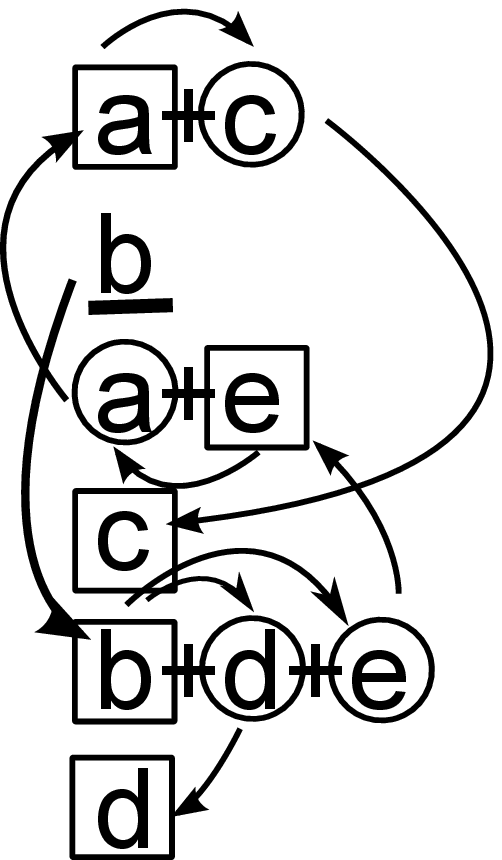}
\label{fig:coh7}
}
\caption[Optional caption for list of figures]{The process of finding the coherent and noncoherent paths to the underlined path in the second subgroup. Coherent paths are put in a circle and noncoherent paths are put in a square.}
\label{fig:Coherent}
\end{figure*}
\else
\begin{figure*}[!t]
\centering
\subfigure[]{
\includegraphics[width = 20mm]{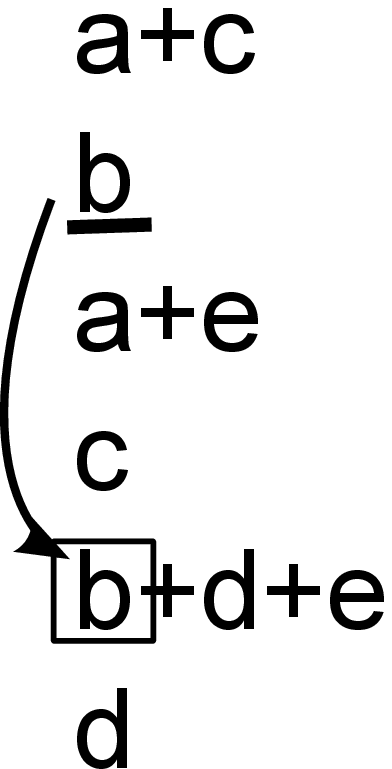}
\label{fig:coh1}
}
\subfigure[]{
\includegraphics[width = 22mm]{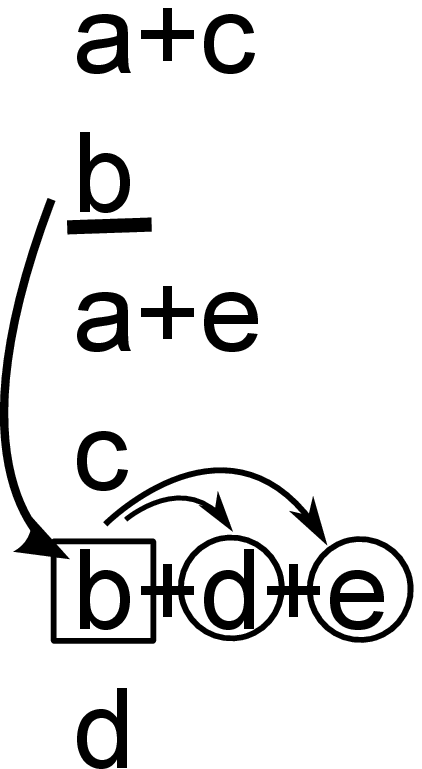}
\label{fig:coh2}
}
\subfigure[]{
\includegraphics[width = 22mm]{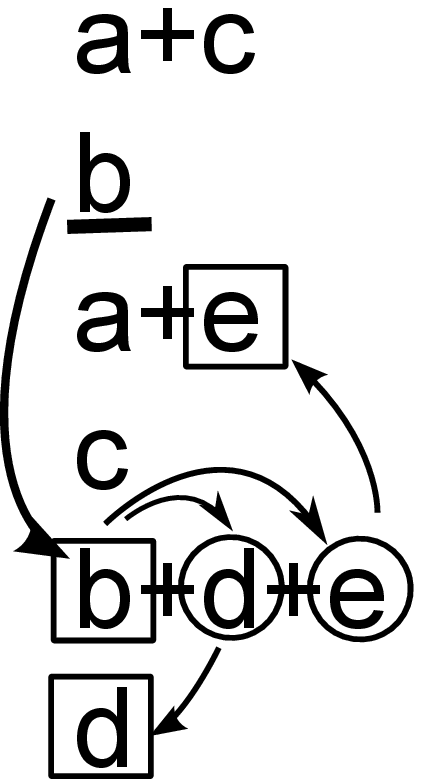}
\label{fig:coh3}
}
\subfigure[]{
\includegraphics[width = 22mm]{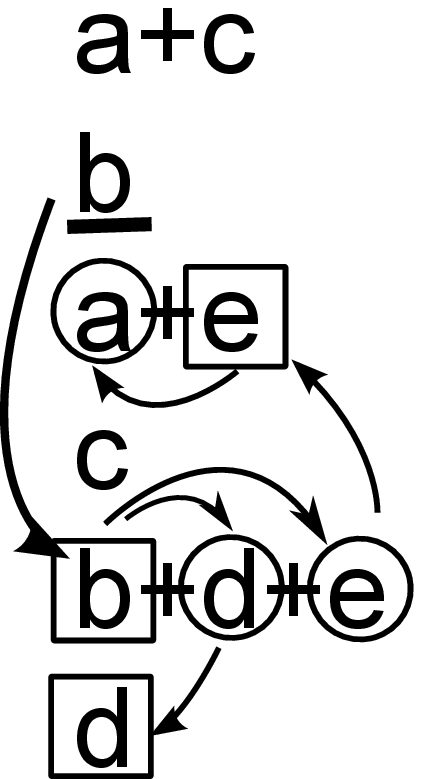}
\label{fig:coh4}
}
\subfigure[]{
\includegraphics[width = 23mm]{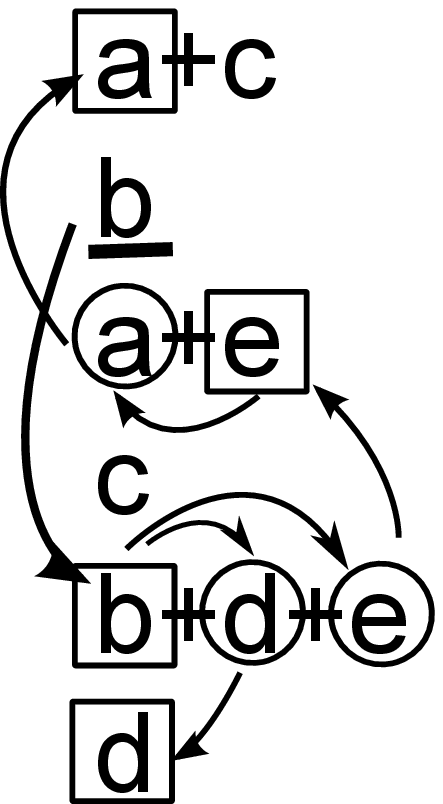}
\label{fig:coh5}
}
\subfigure[]{
\includegraphics[width = 22.7mm]{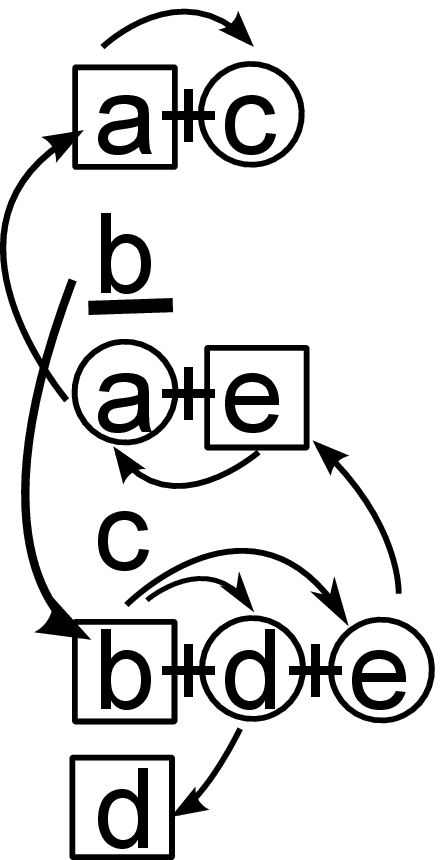}
\label{fig:coh6}
}
\subfigure[]{
\includegraphics[width = 25mm]{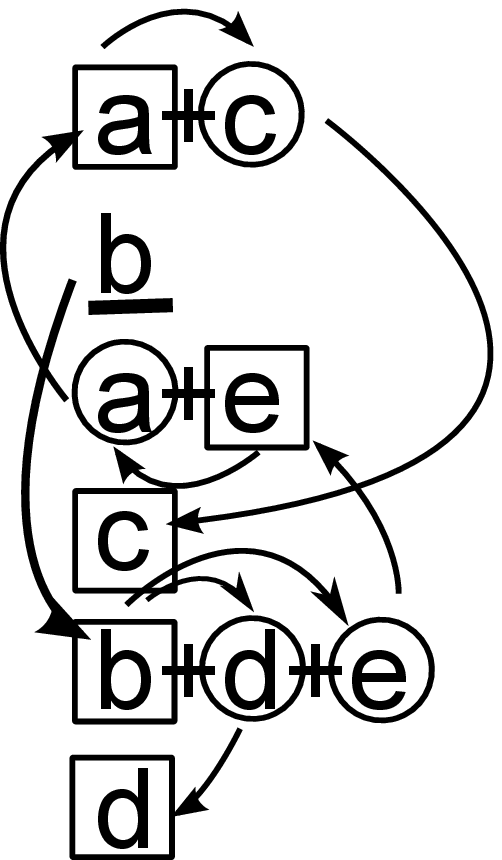}
\label{fig:coh7}
}
\caption[Optional caption for list of figures]{The process of finding the coherent and noncoherent paths to the underlined path in the second subgroup. Coherent paths are put in a circle and noncoherent paths are put in a square.}
\label{fig:Coherent}
\vspace{-5mm}
\end{figure*}
\fi

The received vector of systematic diversity coding for two connection demands looks like
\begin{equation}
\left[\begin{tabular}{c}
$p_1$\\
$p_2$\\
$p_1+p_2$\\
\end{tabular}\right]
\end{equation}
where $p_1$ and $p_2$ are the data signals of two different connection demands. Each symbol on the received vector represents a single path and each data signal is carried with two different paths. The paths carrying the same signal are complementary of each other. If two paths have to be link-disjoint, then they are defined as noncoherent to each other. Assume the path carrying $p_1$ in the first subgroup and the path carrying $p_2$ in the third subgroup fail simultaneously, then the received vector will look like
\begin{equation}
\left[\begin{tabular}{c}
$0$\\
$p_2$\\
$p_1+0$\\
\end{tabular}\right].
\end{equation}
The destination node will still be able to decode symbols $p_1$ and $p_2$. It is clearly seen that diversity coding can tolerate failure of symbols in more than one subgroup. Therefore, the path carrying $p_1$ in the first subgroup and path carrying $p_2$ in the third subgroup can share some of the links. Therefore, they are coherent to each other. Similarly, the path carrying $p_2$ in the second subgroup and the path carrying $p_1$ in the third subgroup can be link-joint. After those relaxations, the solution in Fig.~\ref{fig:disjoint2} is achieved with a modified diversity coding approach. This approach is simpler to keep track of since there are at most $2\times N$ paths for $N$ connection demands. Intuitively, in systematic diversity coding, a path can be link-joint with the paths that are combined with its complementary path. However, to implement those relaxations over nonsystematic codes with an arbitrary number of data signals, a general strategy is needed.
The set of rules that define the general strategy are
\begin{enumerate}
\item[1.] A path has to be link-disjoint (noncoherent) with its complementary path,
\item[2.] A path is coherent with the path that is coded with its complementary path,
\item[3.] A path is noncoherent with the complimentary paths of its coherent paths,
\item[4.] A path is coherent with the paths that are coded with its noncoherent paths.
\end{enumerate}
The logic behind these rules is to make sure that at least one path carrying each data signal survives and any subset of $k$ signals are found within at least $k$ subgroups under any single link failure scenario. It is also important to keep the number of nonzero subgroups greater or equal to $N$ under any failure scenario. The following example visualizes how coherent and noncoherent relationships between paths are found. A valid nonsystematic code is
\begin{equation}
\left[\begin{tabular}{c}
$a+c$\\
$b$\\
$a+e$\\
$c$\\
$b+d+e$\\
$d$\\
\end{tabular}\right]
\end{equation}
with five connection demands. The procedure to find the set of coherent and noncoherent paths of the path carrying $b$ in the second subgroup is shown in Fig.~\ref{fig:Coherent}. In the first step, the complementary path of underlined $b$ is set as a noncoherent path in Fig.~\ref{fig:coh1} following \textit{rule 1}. The coherent paths are put in a circle, whereas noncoherent paths are put in a square. In Fig.~\ref{fig:coh2}, paths that are combined with a noncoherent path are set as coherent paths following \textit{rule 2}. In Fig.~\ref{fig:coh3} and in Fig.~\ref{fig:coh4}, the third and fourth rules of the general strategy are applied, respectively. The process is carried out by following \textit{rule 3} and \textit{rule 4} interchangeably until those rules are no longer applicable. At the end, if there is any nonvisited path in the coding group, it is assumed to be coherent. In that case, the rest of the paths are set as coherent paths to the path of interest.

For example, in Fig.~\ref{fig:Coherent}, assume that the underlined path carrying signal $b$ fails simultaneously with the path carrying signal $a$ in the first subgroup which is noncoherent to itself. If so, the received vector at the destination node becomes
\begin{equation}
\left[\begin{tabular}{c}
$c$\\
$a+e$\\
$c$\\
$b+d+e$\\
$d$\\
\end{tabular}\right].
\end{equation}
This vector clearly violates one of the conditions of decodability because the set of four signals $\{a,e,b,d\}$ are bounded within only three subgroups $\{\{a+e\},\{b+d+e\},\{d\}\}$. Therefore, the resulting decoding vector is not decodable. The other scenarios can also be checked to confirm that simultaneous failures of noncoherent paths impair the survivability, unlike the simultaneous failures of the coherent paths. If more than two paths are supposed to share the same link then each pair of paths must be coherent to each other. To find the coherent and noncoherent set of paths of each path, this process is repeated starting from the path of interest.

We developed an ILP formulation to generate new coding groups based on coherent diversity coding principles. The ILP formulation of this coding structure inherits all of the variables, parameters, objective function and constraints of Section \ref{sec:Nonsys}. The extra variables needed for this ILP formulation are
\begin{itemize}
\item $\phi(i,j) \in \{0,1\}$ : Binary variable, equals 1 iff the path $i$ and path $j$ are coherent, in other words, they can fail simultaneously.
\end{itemize}

The objective function to find a new coding group with the most negative reduced cost is
\begin{equation}
\min \sum_{e\in E}\sum_{s=1}^{2N} t_e(s)\times a_e - \sum_{f\in V} CG_f^{new}\times \pi_f.
\end{equation}
The additional constraints are

\begin{equation}
\theta(i,i-1)=\theta(i-1,i)=1 \hspace{5mm} \forall i :mod(i,2)=0,
\label{eq:coh14}
\end{equation}
\begin{equation}
\theta(i,j) \geq m(i^*,j^*) \hspace{3mm} \forall i,j,
\label{eq:coh12}
\end{equation}
\begin{equation}
\theta(i,k) \geq \theta(i,j) + m(j,k^*)-1 \hspace{5mm} \forall i\neq j\neq k.
\label{eq:coh13}
\end{equation}
\begin{equation}
x_e(i)+x_e(j)+x_{e^*}(i)+x_{e^*}(j) \leq 2- \theta(i,j)  \hspace{5mm} \forall i,j,e
\label{eq:coh15}
\end{equation}
such that link $e$ and link $e^*$ are links of the same span in the opposite directions.
Equation \ref{eq:coh14} makes sure that complimentary paths have to be link-disjoint with each other according to \textit{rule 1}.
Inequality (\ref{eq:coh12}) ensures both \textit{rule 2} and \textit{rule 3} are satisfied. In addition, inequality (\ref{eq:coh13}) ensures that both \textit{rule 3} and \textit{rule 4} are satisfied.
Two paths cannot share a link if they are noncoherent, which is guaranteed by inequality (\ref{eq:coh15}).

\section{Complexity Analysis} \label{sec:complexity}
The column generation method is a very effective technique when used for the implementation of diversity coding since it optimally decomposes the problem into two iterative steps. The alternative coding-based methods, \cite{Kamal2}, \cite{Hover}, and \cite{diversity_journal} usually formulate the coding groups placement problem in a single block. Among those, the diversity coding tree algorithm in \cite{diversity_journal} has significantly fewer variables than the others. On the other hand, in \cite{diversity_glob}, it is shown that implementation of diversity coding with a two step approach is much simpler than the single step approaches. The power of the two step approach is to decompose the bigger main problem into many smaller problems that can be solved in a much shorter time. The complexity of the two step approach also does not depend on the traffic demand matrix. However, the two step approach requires a pre-processing phase where every possible coding group is calculated and enumerated before starting the coding groups placement problem. The number of available coding groups depend on
\begin{equation}
J = \dbinom{|V|-1}{ND-1} + \dbinom{|V|-1}{ND-2} + ... + \dbinom{|V|-1}{1},
\end{equation}
where $ND$ is the nodal degree and $|V|$ is the number of nodes in the network. The number of available coding groups gets exponentially higher as the network size and connectivity increases. It is also costly in terms of memory since it needs to store every possible coding group before starting the coding groups placement problem.

The novelty of the column generation method is to achieve the optimal result without explicitly enumerating all of the possible coding groups. It generates the useful coding groups when they are needed. In the coding groups placement problem, only a very small fraction of all the possible coding groups are placed in the final solution. Therefore, the column generation method needs to generate dramatically fewer coding groups than the two step approach of \cite{diversity_glob}. Table~\ref{table-comparison} highlights the complexity comparison between different optimal  techniques based on LP in terms of the number of integer variables and the number of constraints. The number of continuous variables are negligible compared to the number of integer variables.

\ifCLASSOPTIONonecolumn
\begin{table*}[h!]
\centering
\caption{Complexity Comparisons of the LP Formulations of Different Techniques}
\resizebox{18cm}{!} {
\begin{tabular}{|l|c|c|c|c|c|}
\hline
\multirow{2}{*}{Technique}&\multicolumn{2}{|c|}{Main Problem}&\multicolumn{2}{|c|}{Subproblem}
&\multirow{2}{*}{No. of C.G.} \\ \cline{2-5}
&No. of integer var.&No. of constraints&No. of integer var.&No. of constraints&\\ \hline
MIP in \cite{diversity_journal} &$|N||E|+|N|^2/2$&$3|N||E|/2+|N||V|+7|N|/2$&-&-&1 \\ \hline
ILP in \cite{Kamal2} &$|N|^2/2(|E|+1)+3|N||E|$&$|N|^4/8+...$&-&-&1 \\ \hline
ILP in \cite{Hover}&$|N||E|(|V|+2)+|N|(|N|+2|V|)+...$&$|N||V|(3|E|+|N|+|V|)+...$&-&-&1 \\ \hline
TSA in \cite{diversity_glob}&$J$&$|V|$&$|ND||E|+|ND|^2/2$&$3|ND||E|/2+|ND||V|+...$&$J$ \\ \hline
CGM &$O(|V|)$&$|V|$&$|ND||E|+|ND|^2/2$&$3|ND||E|/2+|ND||V|+...$&$O(|V|)$ \\ \hline
\end{tabular}
}
\label{table-comparison}
\vspace{-3mm}
\end{table*}
\else
\begin{table*}[t]
\centering
\caption{Complexity Comparisons of the LP Formulations of Different Techniques}
\resizebox{18cm}{!} {
\begin{tabular}{|l|c|c|c|c|c|}
\hline
\multirow{2}{*}{Technique}&\multicolumn{2}{|c|}{Main Problem}&\multicolumn{2}{|c|}{Subproblem}
&\multirow{2}{*}{No. of C.G.} \\ \cline{2-5}
&No. of integer var.&No. of constraints&No. of integer var.&No. of constraints&\\ \hline
MIP in \cite{diversity_journal} &$|N||E|+|N|^2/2$&$3|N||E|/2+|N||V|+7|N|/2$&-&-&1 \\ \hline
ILP in \cite{Kamal2} &$|N|^2/2(|E|+1)+3|N||E|$&$|N|^4/8+...$&-&-&1 \\ \hline
ILP in \cite{Hover}&$|N||E|(|V|+2)+|N|(|N|+2|V|)+...$&$|N||V|(3|E|+|N|+|V|)+...$&-&-&1 \\ \hline
TSA &$J$&$|V|$&$|ND||E|+|ND|^2/2$&$3|ND||E|/2+|ND||V|+...$&$J$ \\ \hline
CGM &$O(|V|)$&$|V|$&$(ND)|E|+(ND)^2/2$&$3(ND)|E|/2+(ND)|V|+...$&$O(|V|)$ \\ \hline
\end{tabular}
}
\label{table-comparison}
\vspace{-3mm}
\end{table*}
\fi
TSA is the abbreviation of the two step approach defined in \cite{diversity_glob} and CGM is the proposed column generation method. C.G. corresponds to coding groups. In that table, it is seen that the proposed CGM has dramatically fewer variables and constraints in the main problem. When we assume systematic diversity coding is employed, the subproblem of CGM has fewer variables than competitive techniques. In addition, $ND<<|N|$ and $O(|V|)<<J$. Moreover, in CGM, the complexity of the subproblem and the complexity of the main problem are only linearly added together and multiplied with the average number of useful coding groups. In the first three techniques, the complexity is exponentially dependent on the number of unit traffic demands and the network size. In TSA, the complexity is dependent on the number of candidate coding groups, which exponentially increases with $|N|$ and $ND$. In addition, TSA works on much more coding groups than CGM does. As a result, the proposed CGM is much simpler and scalable than competitive techniques. Therefore, it can implement diversity coding over very large arbitrary networks.
The simplicity of the CGM is also reflected in the next section in terms of runtime of different algorithms.
In \cite{diversity_journal}, it is explained that adopting single destination diversity coding enables near-hitless recovery and simplifies design complexity significantly. Therefore, in this paper, single destination diversity coding is adopted. 
\section{Theoretical Lower Bound}
In this section, we look into the theoretical limits of the capacity requirements of single destination coding-based
recovery techniques. The derived lower bounds will be helpful to understand the space of improvement over
our proposed techniques, which already can be implemented on very large real networks. Advancement in coding
techniques usually results in much higher design complexity which prevents to apply them on real networks,
such as \cite{rouy}. Therefore, we need to find out the extent of incentive to developed more advanced coding techniques in terms of capacity efficiency.

A coding-based recovery technique can recover from single link failures instantaneously as long as it can support
the traffic demand even if any of the edges are removed from the network \cite{rouy}. In \cite{rouy}, it is mathematically stated using max-flow min-cut theorem \cite{GroverBook} as
\begin{equation}
\min_{C}\left[ \sum_{e\in C(E)}c(e) - \max_{e\in C(e)}c(e) \right]\geq h,
\label{eq:tlb1}
\end{equation}
where $C$ is a cut that disconnects the source node from the destination, $C(E)$ is the set of links
in cut $C$, $c(e)$ is the capacity put on link $e$, and $h$ is the total demand. In \cite{rouy}, there is a single source and destination node. However, we assume multiple source nodes and a single destination node. Therefore, we need to take the partial cuts, which disconnects a subset of source nodes from the destination, into account. In addition, we assume an edge includes two opposite directional links which fail simultaneously. The inequality (\ref{eq:tlb1}) is modified as
\begin{equation}
\min_{C}\left[ \sum_{e\in C(E)}c(e) - \max_{e\in C(e)}c(e) \right]\geq \sum_{s_i\in V(C)}t(s_i),
\label{eq:tlb2}
\end{equation}
where $V(C)$ is the set of source nodes disconnected due to cut $C$ and $t(s_i)$ is the demand of source node $s_i$.

We have benefited from an ILP formulation to find the lower bound of capacity requirements over arbitrary networks. This formulation finds the minimum capacity placements that satisfy the cut capacity criterions stated by inequality (\ref{eq:tlb2}). Inequality (\ref{eq:tlb2}) includes nonlinear operations therefore, we need to apply a set of conversions to make it linear, such as

\begin{equation}
\sum_{e\in C(E)}c(e) - \max_{e\in C(e)}c(e) \geq \sum_{s_i\in V(C)} t(s_i), \hspace{5mm} \forall C.
\label{eq:tlb3}
\end{equation}
After one more conversion, inequality (\ref{eq:tlb3}) becomes
\begin{equation}
\sum_{e\in C(E)}c(e) - c(f) \geq \sum_{s_i\in V(C)}t(s_i), \hspace{5mm} \forall f\in C(E), \forall C.
\label{eq:tlb4}
\end{equation}
In the final step, the only set of constraints of the ILP formulation is
\ifCLASSOPTIONonecolumn
\begin{equation}
\sum_{e\in C(E)}(c(m_1)+c(m_2)) - c(l_1) - c(l_2)\geq \sum_{s_i\in V(C)}t(s_i), \hspace{5mm} \forall l_1,l_2 \in f,\forall f\in C(E), \forall C,
\label{eq:tlb5}
\end{equation}
\else
\begin{eqnarray} \nonumber
\sum_{e\in C(E)}(c(m_1)+c(m_2)) - c(l_1) - c(l_2)\geq \sum_{s_i\in V(C)}t(s_i),\\
\forall l_1,l_2 \in f,\forall f\in C(E), \forall C,
\label{eq:tlb5}
\end{eqnarray}
\fi
where $c(i)$ is an integer variable that keeps the required capacity over link $i$. The symbols $m_1$ and $m_2$ are opposite directional links over edge $e$.
The objective function of the ILP formulation is
\begin{equation}
\min \sum_{l\in E} c(l) \times a_l,
\label{eq:tlb6}
\end{equation}
where $a_l$ is the length of link $l$.

The scalability of this operation is questionable since the number of all possible cuts is proportional to
\begin{equation}
\dbinom{|E|}{1} + \dbinom{|E|}{2} + ... + \dbinom{|E|}{|E|}.
\end{equation}
Even for a small network with 20 edges, the total number of cuts exceeds millions. Moreover, for each cut with a size $|C(E)|$, we require $|C(E)|$ inequalities. Therefore, it is very difficult to derive the tightest lower bound by investigating all possible cuts. Therefore, we opt to look into cuts including limited number of edges that gives an approximate lower bound.    
\section{Simulation Results}
\label{sec:Results}

In this section, we present various simulation results to investigate the performance of the novel design algorithm and the new coding structure differentially. The first test network is the NSFNET network, which is depicted in Fig.~\ref{NSFNET network}. The number next to the nodes are the index of those nodes and the numbers next to the edges are the length of those edges. The traffic matrix of the NSFNET network consists of 3000 random unit-sized demands, which are chosen using a realistic gravity-based model \cite{zrdg}. Each node in the NSFNET network represents a U.S. metropolitan area and their population is proportional to the weight of each node in the connection demand selection process. In this network, we simulated TSA from \cite{diversity_glob}, {\em p}-cycle protection \cite{pcycle_col}, diversity coding tree from \cite{diversity_journal} and the proposed CGM. CPLEX 12.2 is used for the simulations. We also adopted different coding structures for TSA and CGM. There are three different tables that present the simulation results of this network. In Table~\ref{table-CapEff2}, the performance metrics are the total capacity (TC) and the runtime. The first technique in this table is the diversity coding tree algorithm. TSA-SDC refers to the two-step approach implementing systematic diversity coding, whereas TSA-NSDC means TSA for nonsystematic diversity coding. CGM-SDC, CGM-NSDC, and CGM-CDC correspond to the CGM implementing systematic diversity coding, nonsystematic diversity coding, and coherent diversity coding. It is noted that these three algorithms are implemented sequentially. The coding groups (columns) generated by CGM-SDC are inherited by CGM-NSDC. Likewise, CGM-CDC inherits the coding groups generated by CGM-NSDC.
The {\em p}-cycle algorithm is taken from \cite{pcycle_col}, which is also based on column generation.

Table~\ref{table-CapEff2} presents various trade-offs between protection techniques. First of all, the coding-based techniques are able to offer near-hitless recovery. Their restoration speed is at least two orders of magnitude higher than that of {\em p}-cycle protection \cite{diversity_journal}. On the other hand, {\em p}-cycle protection has higher capacity efficiency than the tested coding-based methods. As it is seen, the diversity coding tree algorithm has the highest complexity which keeps it from achieving optimal results even though it implements the same systematic diversity coding like TSA-SDC and CGM-SDC do. The proposed CGM is more scalable than the diversity coding tree algorithm and TSA, as seen from the runtime column. In both TSA and CGM, nonsystematic diversity coding is more capacity efficient than systematic diversity coding. In addition, proposed coherent diversity coding is the most capacity efficient among coding-based method. However, the increase in capacity efficiency is negligible compared to the savings in runtime. Network designers can opt to carry out the implementations of CGM-NSDC and CGM-CDC after the implementation of CGM-SDC. We believe that CGM-SDC is the most efficient coding-based technique in terms of restoration speed, capacity efficiency, and design complexity.

\ifCLASSOPTIONonecolumn
\begin{figure}[!h]
\centering
\includegraphics[bb = 45 167 707 492,width = 85mm,height = 43mm,clip=true]{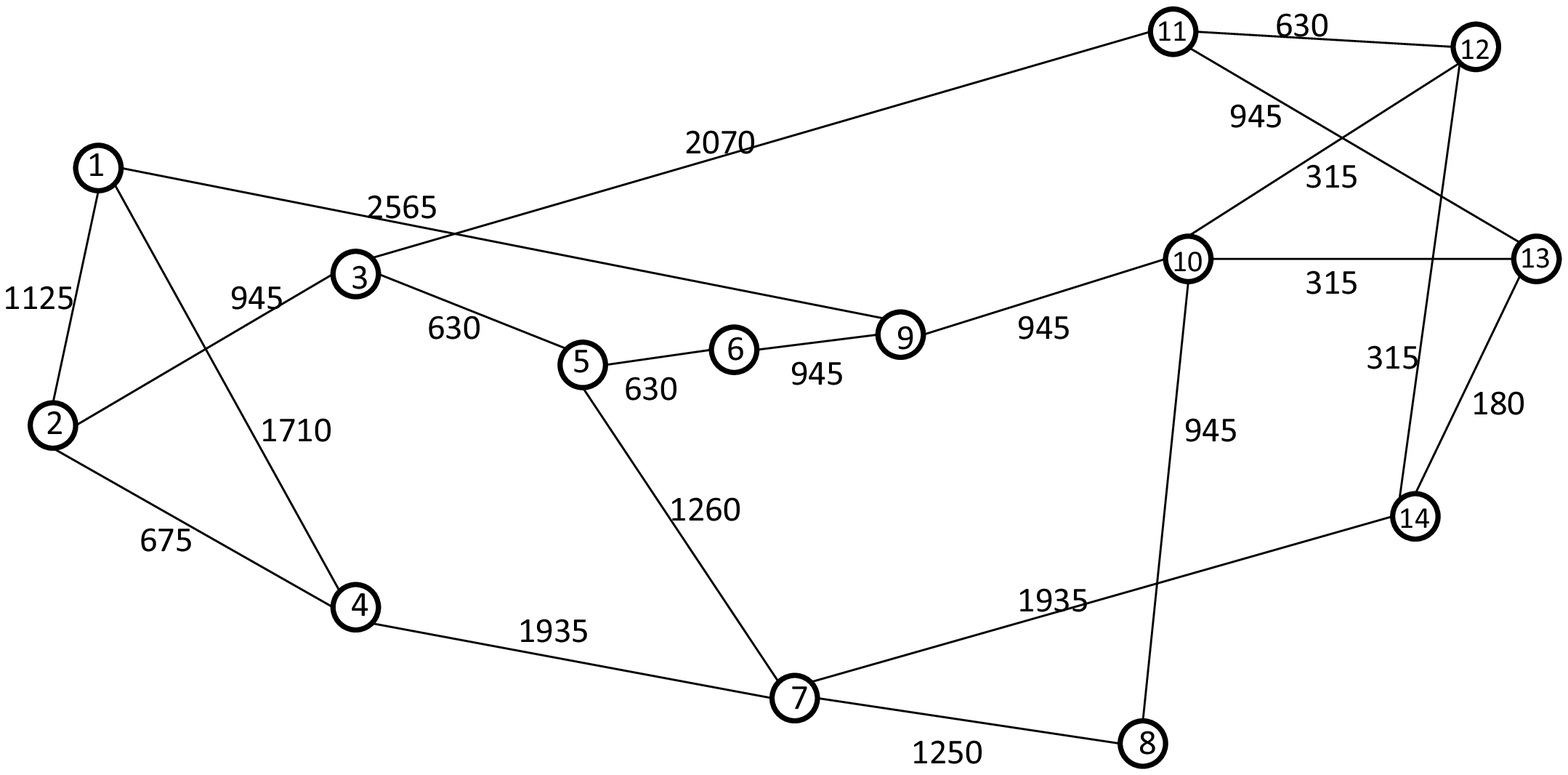}
\caption{NSFNET network.}
\label{NSFNET network}
\end{figure}
\else
\begin{figure}[!b]
\centering
\includegraphics[bb = 45 167 707 492, width = 85mm,height = 43mm,clip=true]{NSFNET.eps}
\caption{NSFNET network.}
\label{NSFNET network}
\end{figure}
\fi

\ifCLASSOPTIONonecolumn
\begin{table*}[h!]
\centering
\caption{Cost and Runtime Comparison between Different Techniques}
\begin{tabular}{|l|c|c|}
\hline
Protection Technique&Total Cost&Runtime \\ \hline
Diversity Coding Tree &16880400&$\approx$ 6 hours \\ \hline
TSA-SDC&15788730&$\approx$ 6 minutes \\ \hline
TSA-NSDC&15746690&$\approx$ 9 minutes \\ \hline
CGM-SDC &15793170&$\approx$ 10 seconds \\ \hline
CGM-NSDC&15746690&$\approx$ 5 minutes \\ \hline
CGM-CDC&15678770&$\approx$ 1 hour \\ \hline
{\em P}-cycle algorithm &14814350 &$\approx$ 3 minutes \\ \hline
\end{tabular}
\label{table-CapEff2}
\end{table*}
\else
\begin{table*}[t]
\centering
\caption{Cost and Runtime Comparison between Different Techniques}
\begin{tabular}{|l|c|c|c|}
\hline
Protection Technique&Total Cost&Runtime \\ \hline
Diversity Coding Tree &16880400&$\approx$ 6 hours \\ \hline
TSA-SDC&15788730&$\approx$ 6 minutes \\ \hline
TSA-NSDC&15746690&$\approx$ 9 minutes \\ \hline
CGM-SDC &15793170&$\approx$ 10 seconds \\ \hline
CGM-NSDC&15746690&$\approx$ 5 minutes \\ \hline
CGM-CDC&15678770&$\approx$ 1 hour \\ \hline
{\em P}-cycle algorithm &14814350 &$\approx$ 3 minutes \\ \hline
\end{tabular}
\label{table-CapEff2}
\end{table*}
\fi

In Table~\ref{table-CapEff}, the performance of the nonsystematic and coherent diversity coding is shown compared to the systematic diversity coding and the theoretical lower bound with a breakdown over the nodes. It must be noted that the lower bound is not the tightest bound due to the exponential complexity of calculating it. The full and partial cuts with at most 5 edges are taken into account. For the tightest bound, all possible cuts should be investigated. The performance metric is the TC to route and protect the connection demands. The goal is to measure the decrease in TC due to the introduction of nonsystematic and coherent diversity coding. As expected, nonsystematic diversity coding performs better than systematic diversity coding, whereas coherent diversity coding performs best of all. As mentioned before, the improvement due to introduction of advanced coding techniques is limited over all different destination node scenarios. The difference between total capacity required by coherent diversity coding and the theoretical lower bound varies between 0-11\% depending on the destination node. On average, coherent diversity coding requires 7.1\% extra capacity than the theoretical lower bound, which is expected to get smaller if a tighter lower bound can be achieved bearing more complex calculations.

\ifCLASSOPTIONonecolumn
\begin{table*}[h!]
\centering
\caption{NSFNET, TC Results for Each Destination Node}
\begin{tabular}{|l|c|c|c|c|}
\hline
Destination Node & CGM-SDC & CGM-NSDC & CGM-CDC & Theoretical Lower Bound \\ \hline
Node 1&762550&762550&762550&753450 \\ \hline
Node 2&1301320&1301320&1301320&1215250 \\ \hline
Node 3&211500&211500&211500&199350 \\ \hline
Node 4&3913950&3913950&3913950&3709300 \\ \hline
Node 5&455100&455100&455100&415600 \\ \hline
Node 6&128150&128150&128150&128150 \\ \hline
Node 7&1119200&1072720&1072720&1010350 \\ \hline
Node 8&602000&602000&595700&536300 \\ \hline
Node 9&1595700&1595700&1595700&1484850 \\ \hline
Node 10&1177230&1172970&1172970&1054550 \\ \hline
Node 11&573650&573650&573650&546650 \\ \hline
Node 12&2293720&2293720&2264700&2074550 \\ \hline
Node 13&1150350&1150350&1130350&1036700 \\ \hline
Node 14&508750&508750&496150&473650 \\ \hline
Total&15793170&15746690&15678770&14638700 \\ \hline
\end{tabular}
\label{table-CapEff}
\end{table*}
\else
\begin{table*}[t]
\centering
\caption{NSFNET, TC Results for Each Destination Node}
\begin{tabular}{|l|c|c|c|c|}
\hline
Destination Node & CGM-SDC & CGM-NSDC & CGM-CDC & Theoretical Lower Bound \\ \hline
Node 1&762550&762550&762550&753450 \\ \hline
Node 2&1301320&1301320&1301320&1215250 \\ \hline
Node 3&211500&211500&211500&199350 \\ \hline
Node 4&3913950&3913950&3913950&3709300 \\ \hline
Node 5&455100&455100&455100&415600 \\ \hline
Node 6&128150&128150&128150&128150 \\ \hline
Node 7&1119200&1072720&1072720&1010350 \\ \hline
Node 8&602000&602000&595700&536300 \\ \hline
Node 9&1595700&1595700&1595700&1484850 \\ \hline
Node 10&1177230&1172970&1172970&1054550 \\ \hline
Node 11&573650&573650&573650&546650 \\ \hline
Node 12&2293720&2293720&2264700&2074550 \\ \hline
Node 13&1150350&1150350&1130350&1036700 \\ \hline
Node 14&508750&508750&496150&473650 \\ \hline
Total&15793170&15746690&15678770&14638700 \\ \hline
\end{tabular}
\label{table-CapEff}
\end{table*}
\fi

In Table~\ref{table-Granu}, the effect of the traffic granularity is investigated over the total cost, the LP lower bound of the integer solution and the optimality gap between the ILP solution and the LP lower bound. We input three different traffic scenarios. In the first scenario, there are 300 unit connection demands created by the gravity-based model. In the second scenario, each traffic demand is divided into 10 smaller unit connection demands creating 3000 connection demands. In the final scenario, 30000 connection demands are created by doing the same operation again. It is seen from the results, the optimality gap decreases as the granularity of the connection demands decreases. Optimality gap converges to zero fast. The fact that column generation is implemented over LP increases the simulation speed significantly but does not deteriorate the performance.

\ifCLASSOPTIONonecolumn
\begin{table*}[h!]
\centering
\caption{The effect of granularity on network optimization}
\begin{tabular}{|l|c|c|c|c|}
\hline
Protection Technique&No. of connection demands&Total Cost&LP Bound&Optimality Gap \\ \hline
CGM-SDC&300  &1602780&1578407&$\approx$ 1.47\% \\ \hline
CGM-SDC&3000 &1579355&1578407&$\approx$ 0.06\% \\ \hline
CGM-SDC&30000&1579317&1578407&$\approx$ 0.06\% \\ \hline
\end{tabular}
\label{table-Granu}
\end{table*}
\else
\begin{table*}[t]
\centering
\caption{The effect of granularity on network optimization}
\begin{tabular}{|l|c|c|c|c|}
\hline
Protection Technique&No. of connection demands&Total Cost&LP Bound&Optimality Gap \\ \hline
CGM-SDC&300  &1602780&1578407&$\approx$ 1.47\% \\ \hline
CGM-SDC&3000 &1579355&1578407&$\approx$ 0.06\% \\ \hline
CGM-SDC&30000&1579317&1578407&$\approx$ 0.06\% \\ \hline
\end{tabular}
\label{table-Granu}
\end{table*}
\fi


\ifCLASSOPTIONonecolumn
\begin{table*}[h!]
\centering
\caption{Comparative performance of the new algorithms in U.S. long-distance network}
\begin{tabular}{|l|c|c|c|}
\hline
Protection Technique&SCaP&Runtime&No. of Coding Groups \\ \hline
TSA-SDC&105.6\%&$\approx$ 3 hours&31464 \\ \hline
CGM-SDC&105.6\%&$\approx$ 2 minutes&61 \\ \hline
CGM-NSDC&105.5\%&$\approx$ 2 hours&72 \\ \hline
CGM-CDC&93.4\%&$\approx$ 9 hours&79 \\ \hline
{\em P}-cycle algorithm &107.0\%&$\approx$ 2.5 hours&32 ({\em p}-cycles) \\ \hline
\end{tabular}
\label{table-Large}
\end{table*}
\else
\begin{table*}[t]
\centering
\caption{Comparative performance of the new algorithms in U.S. long-distance network}
\begin{tabular}{|l|c|c|c|}
\hline
Protection Technique&SCaP&Runtime&No. of Coding Groups \\ \hline
TSA-SDC&105.6\%&$\approx$ 3 hours&31464 \\ \hline
CGM-SDC&105.6\%&$\approx$ 2 minutes&61 \\ \hline
CGM-NSDC&105.5\%&$\approx$ 2 hours&72 \\ \hline
CGM-CDC&93.4\%&$\approx$ 9 hours&79 \\ \hline
{\em P}-cycle algorithm &107.0\%&$\approx$ 2.5 hours&32 ({\em p}-cycles) \\ \hline
\end{tabular}
\label{table-Large}
\end{table*}
\fi
The second test network is the U.S. long-distance network, taken from \cite{XM99}, which is shown in Fig.~\ref{fig:USlong}. The traffic matrix is created using a gravity-based model \cite{zrdg}. In total, there are 23,204 static unit connection demands. This setup is chosen in order to observe the performance of the new design algorithm in a large realistic network with a dense traffic scenario. We compare the performance of CGM with TSA and {\em p}-cycle algorithm from \cite{pcycle_col} in terms of spare capacity percentage (SCaP) defined in \cite{diversity}. The other coding-based recovery design algorithms are too complex to implement in this setup. The results are presented in Table~\ref{table-Large}.
\ifCLASSOPTIONonecolumn
\begin{figure}[h!]
\centering
\includegraphics[width=80mm]{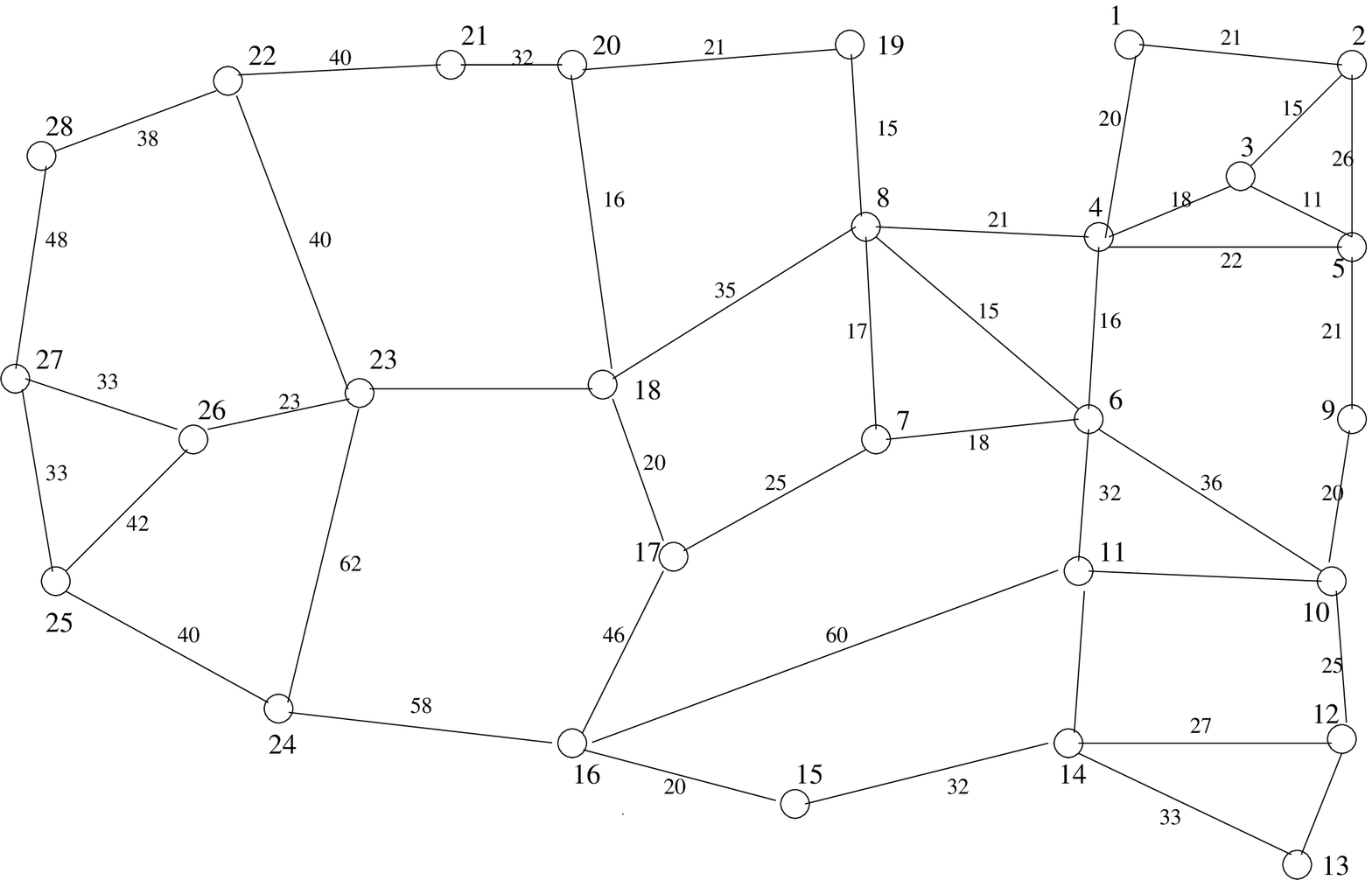}
\caption{U.S. long-distance network.}
\label{fig:USlong}
\end{figure}
\else
\begin{figure}[!t]
\centering
\includegraphics[width=80mm]{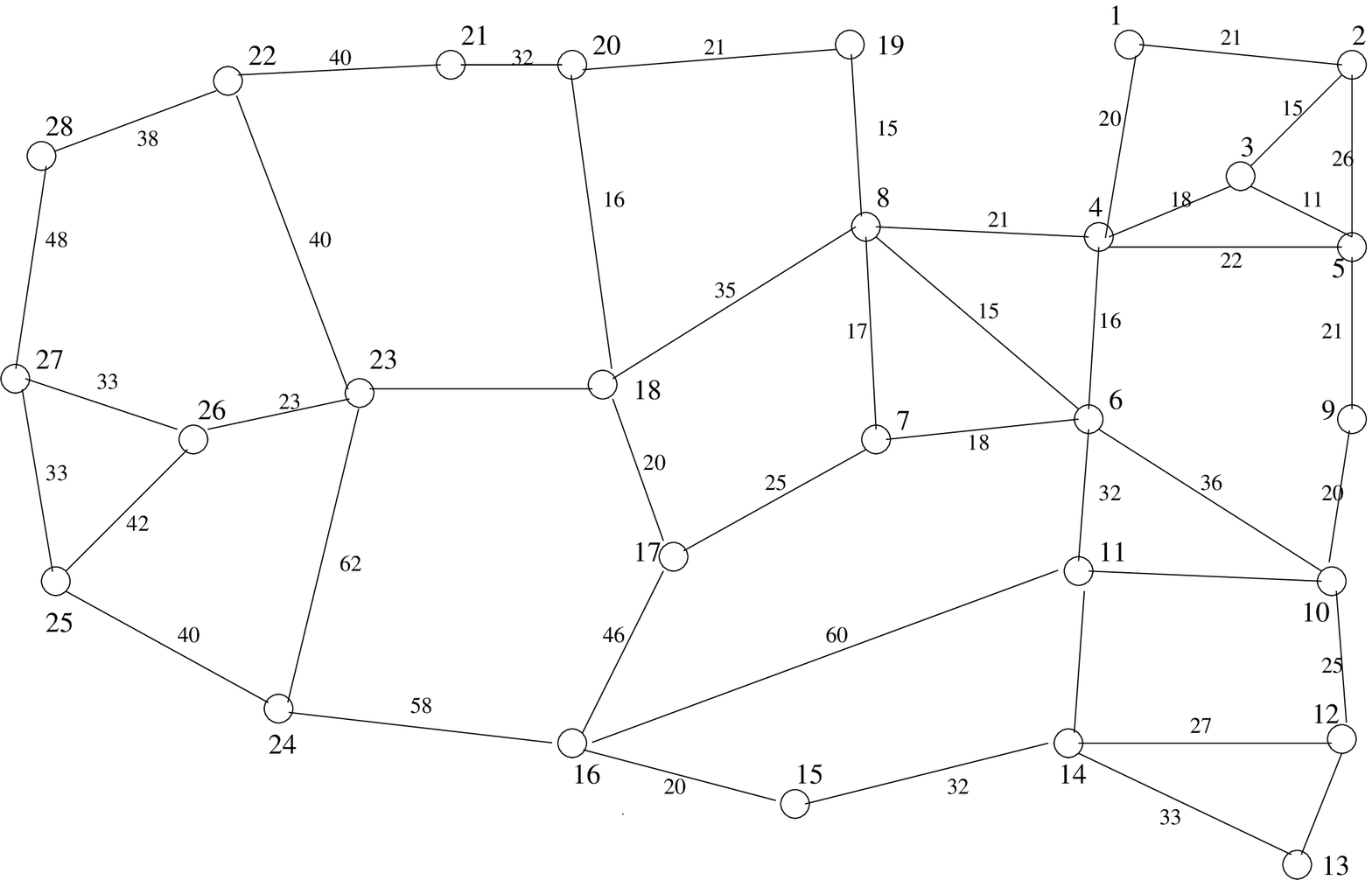}
\caption{U.S. long-distance network.}
\label{fig:USlong}
\end{figure}
\fi

As seen from the results, the proposed design algorithm can achieve optimal results with different versions of diversity coding even in a large realistic network with a dense traffic scenario. Proposed coherent diversity coding technique performs best compared to other coding-based recovery techniques at the expense of higher complexity. The increase in capacity efficiency due to the advanced coding technique is more significant than it is in the NSFNET network. The implementation of systematic diversity coding with the proposed CGM is highly scalable since its runtime does not increase as much as others when the network size gets bigger. The TSA approach is not as scalable as CGM since the number of candidate paths in TSA increases exponentially with the nodal degree and the number of nodes, whereas the number of candidate paths in CGM increases linearly with the number of nodes. The SCaP result of the new technique is better than that of the column generation based {\em p}-cycle algorithm. It should be noted that, {\em p}-cycle algorithm carries out Spare Capacity Placement (SCP) \cite{GroverBook} due to high complexity, whereas the proposed algorithm carries out Joint Capacity Placement (JCP) \cite{GroverBook}. Even with that adjustment, the proposed CGM is simpler than the {\em p}-cycle algorithm.

The third network is the long-distance network of France with 43 nodes and 142 unidirectional links taken from \cite{doucette}. It is depicted in Fig.~\ref{fig:France}. There are a total number of 4518318 unit connection demands. The traffic scenario is created following the same gravity-based model. The reason to select this network is to test the performance of CGM in very large realistic networks. Therefore, we only simulate CGM-SDC to investigate the runtime performance of the column generation method without extra complexity due to the advanced coding structure. It is compared to 1+1 APS. We also break down the results in terms of the nodal degree of the nodes to see the effect of the nodal degree on both capacity efficiency and runtime. The results are presented in Table~\ref{table-France}. The runtime of 1+1 APS is equal to 1 minute.

\ifCLASSOPTIONonecolumn
\begin{table*}[h!]
\centering
\caption{SCaP performance of the new algorithm with respect to the nodal degree}
\begin{tabular}{|l|c|c|c|c|}
\hline
Nodal Degree&CGM-SDC (SCaP)&1+1 APS (SCaP)&Runtime of CGM-SDC&Sample Size \\ \hline
2 links&155.3\%&155.3\%&$\approx$ 2 minutes&12 nodes \\ \hline
3 links&125.5\%&149.4\%&$\approx$ 16 minutes&13 nodes \\ \hline
4 links&106.7\%&140.6\%&$\approx$ 39 minutes&14 nodes \\ \hline
5 links&146.4\%&184.5\%&$\approx$ 26 minutes&2 nodes \\ \hline
6 links&89.5\%&126.5\%&$\approx$ 85 minutes&1 node \\ \hline
7 links&86.6\%&136.6\%&$\approx$ 53 minutes&1 node \\ \hline
Total&105.7\%&141.0\%&$\approx$ 85 minutes& 43 nodes \\ \hline
\end{tabular}
\label{table-France}
\end{table*}
\else
\begin{table*}[t]
\centering
\caption{SCaP performance of the new algorithm with respect to the nodal degree}
\begin{tabular}{|l|c|c|c|c|}
\hline
Nodal Degree&CGM-SDC (SCaP)&1+1 APS (SCaP)&Runtime of CGM-SDC&Sample Size \\ \hline
2 links&155.3\%&155.3\%&$\approx$ 2 minutes&12 nodes \\ \hline
3 links&125.5\%&149.4\%&$\approx$ 16 minutes&13 nodes \\ \hline
4 links&106.7\%&140.6\%&$\approx$ 39 minutes&14 nodes \\ \hline
5 links&146.4\%&184.5\%&$\approx$ 26 minutes&2 nodes \\ \hline
6 links&89.5\%&126.5\%&$\approx$ 85 minutes&1 node \\ \hline
7 links&86.6\%&136.6&$\approx$ 53 minutes&1 node \\ \hline
Total&105.7\%&141.0\%&$\approx$ 85 minutes& 43 nodes \\ \hline
\end{tabular}
\label{table-France}
\end{table*}
\fi

\ifCLASSOPTIONonecolumn
\begin{figure}[h!]
\centering
\includegraphics[width=120mm]{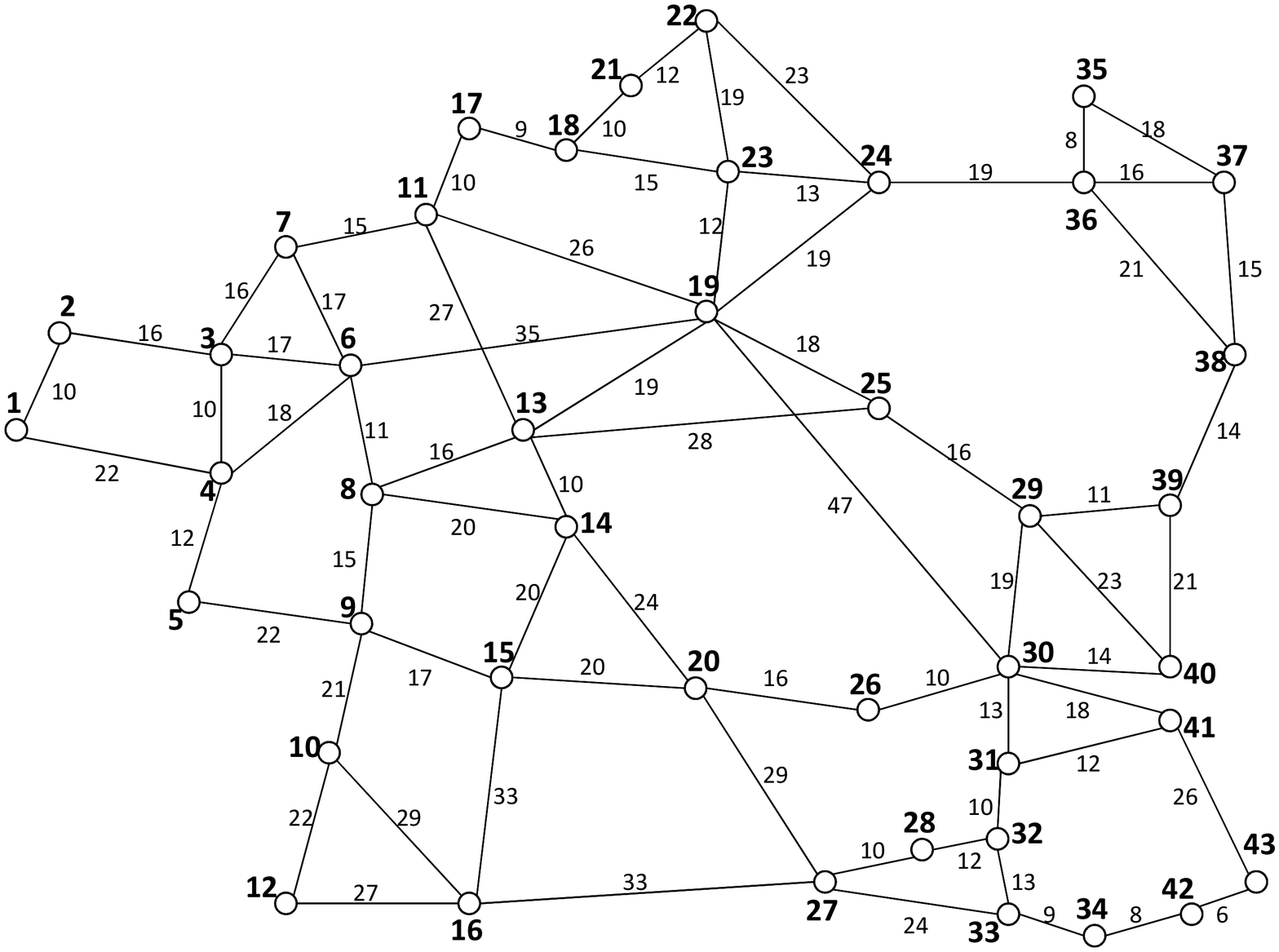}
\caption{Long-distance network of France.}
\label{fig:France}
\end{figure}
\else
\begin{figure}[!t]
\centering
\includegraphics[width=80mm]{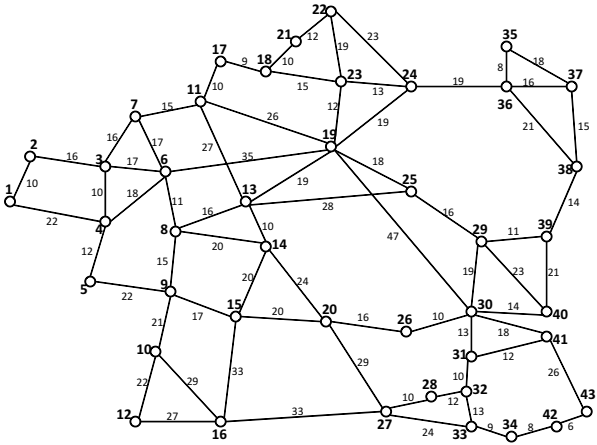}
\caption{Long-distance network of France.}
\label{fig:France}
\end{figure}
\fi

CGM can achieve the optimal result in such a large network with over four million unit demands. The capacity efficiency of CGM-SDC improves as the nodal degree increases with the exception of nodal degree being equal to 5. It may be seen as an exception due to the small sample size. According to the Table~\ref{table-France}, there is a trade-off between the runtime and the capacity improvement over 1+1 APS. When the nodal degree increases, the SCaP improvement of CGM-SDC over 1+1 APS increases at the expense of increased runtime of CGM-SDC with some exceptions due to the small sample size. When the nodal degree is equal to 2, diversity coding acts the same as 1+1 APS as we mentioned before.

\section{Conclusion}

In this paper, we introduced an advanced version of diversity coding and an optimal and simple design algorithm to achieve near instantaneous recovery with higher capacity efficiency. The proposed coherent diversity coding method employs nonsystematic coding, which enables all paths to be encoded, and relaxes the link-disjointness criterion between paths to cope with the low nodal degree in the network. The code is developed with the objective of minimum capacity. The design algorithm consists of two parts, namely a main problem and subproblem. These two advanced techniques combined achieve results with higher capacity efficiency in a much shorter amount of time in relatively large networks. The advantages of both techniques are shown with examples and simulation results.

The new design framework is based on column-generation method and consist of two parts, a main problem where the traffic demands are met with the available coding groups and the subproblem where new useful coding groups are generated at each iteration. The main problem starts with a set of dummy coding groups and inputs new coding groups at each iteration. The subproblem creates a new coding group depending on the information coming from the main problem. The iterations are terminated when a new useful coding group cannot be found. The main problem is formulated as LP throughout the iteration process. At the end, the main problem is solved via ILP which creates a very small optimality gap. We have formulated the subproblem different for different coding techniques based on either ILP or MIP. There is a complexity versus capacity efficiency trade off in formulating the subproblem. The main problem consists of only $|V|-1$ constraints. It finds and places the optimal coding group combinations to match the traffic demands, which takes sub-ms to run. The new algorithm can be implemented over networks with arbitrary topology and it can achieve optimal results in very large arbitrary networks for arbitrary traffic scenarios.

We ran various sets of simulations to investigate the performance of the new coding structure and the new design algorithm differentially. The coherent diversity coding has a higher capacity efficiency then both the nonsystematic and systematic diversity coding. The improvement is very small in some networks but is more significant in other networks. The most important observation of the paper is how the new column generation-based design method simplifies implementation of coding-based recovery techniques in very large arbitrary networks. The new technique can find optimal solutions in a much shorter time then the competitive techniques. The complexity of the new technique is also more scalable then the competitive techniques depending on the network size, the size of the traffic demands, and the nodal degree of the nodes in the network.

%
%

\vspace{-1.5mm}
\bibliographystyle{IEEEtran}
\bibliography{IEEEabrv,bibliography/ISIT}
\end{document}